\documentclass[nomathfonts]{aipproc}
\layoutstyle{6s}

\usepackage{xspace,epic,eepic,graphicx}
\graphicspath{{figures/}}

\title[Bayesian source separation]
      {Bayesian source separation with mixture of Gaussians prior 
       for sources and Gaussian prior for mixture coefficients}  
\author{Hichem Snoussi}
  {
  address = {Laboratoire des Signaux et Syst\`emes (L2S), \linebreak   
             Sup\'elec, Plateau de Moulon,  
				 91192 Gif-sur-Yvette Cedex, France}
  ,email   = snoussi@lss.supelec.fr
  }
\author{Ali Mohammad-Djafari}
  {
  address = {Laboratoire des Signaux et Syst\`emes (L2S), \linebreak  
             Sup\'elec, Plateau de Moulon, 
				 91192 Gif-sur-Yvette Cedex, France}
  ,email   = djafari@lss.supelec.fr
  }

\input{inputs/alphabet}       
\input{inputs/abrege}         
\def\bigcro#1{\bigl[#1\bigr]}

\def\Exp#1{\exp\bigcro{#1}}

\newsavebox{\fminibox}
\newlength{\fminilength}
\newenvironment{fminipage}[1][\linewidth]
  {\setlength{\fminilength}{#1}
   \begin{lrbox}{\fminibox}\begin{minipage}{\fminilength}}
  {\end{minipage}\end{lrbox}\noindent\fbox{\usebox{\fminibox}}}

  \def\+{^\dagger}

\def\nequiv{\not\kern-.05em\equiv}
\def\egal{\kern-.5em=\kern-.5em}        
\def\propt{\kern-.2em\propto\kern-.2em} 
\def\wh#1{\widehat{#1}}                 

\def\argmax{\mathop{\mathrm{arg\,max}}} 

\def\intdouble{\int\kern-0.3em\int}
\def\inttriple{\int\kern-0.3em\int\kern-0.3em\int}

\def\rond#1{\overset{\kern-0.33em~_\circ}{#1}}
\def\rondit[#1]#2{\overset{\kern#1~_\circ}{#2}}

\def\babs{\begin{abstract}}             \def\eabs{\end{abstract}}
\def\barr{\begin{array}}                \def\earr{\end{array}}
\def\bcc{\begin{center}}                \def\ecc{\end{center}}

\def\bdes{\begin{description}}          \def\edes{\end{description}}
\def\bdoc{
\begin{document}}             \def\edoc{\end{document}}
\def\ben{\begin{enumerate}}             \def\een{\end{enumerate}}
\def\beqn{\begin{eqnarray}}             \def\eeqn{\end{eqnarray}}
\def\beqnl#1{\beqn\label{#1}}           \def\eeqnl#1{\label{#1}\eeqn}
\def\beqnx{\begin{eqnarray*}}           \def\eeqnx{\end{eqnarray*}}
\def\bseqn{\begin{subeqnarray}}         \def\eseqn{\end{subeqnarray}}
\def\beq#1\eeq{\begin{equation}#1\end{equation}}
\def\bal#1\eal{\begin{align}#1\end{align}}
\def\balx#1\ealx{\begin{align*}#1\end{align*}}
\def\beqx{$$}                           \def\eeqx{$$}
\def\bfig{\protect\begin{figure}}       \def\efig{\protect\end{figure}}
\def\bfigx{\protect\begin{figure*}}     \def\efigx{\protect\end{figure*}}
\def\bfigt{\protect\begin{figurette}}   \def\efigt{\protect\end{figurette}}
\def\bfl{\begin{flushleft}}             \def\efl{\end{flushleft}}
\def\bfr{\begin{flushright}}            \def\efr{\end{flushright}}
\def\bit{\begin{itemize}}               \def\eit{\end{itemize}}
\def\bmi{\begin{minipage}}              \def\emi{\end{minipage}}
\def\bfmi{\begin{fminipage}}            \def\efmi{\end{fminipage}}
\def\bpic{\begin{picture}}              \def\epic{\end{picture}}
\def\bqu{\begin{quote}}                 \def\equ{\end{quote}}
\def\bqun{\begin{quotation}}            \def\equn{\end{quotation}}
\def\bsl{\begin{slide}}                 \def\esl{\end{slide}}
\def\btabb{\begin{tabbing}}             \def\etabb{\end{tabbing}}
\def\btabl{\begin{table}}               \def\etabl{\end{table}}
\def\btablx{\begin{table*}}             \def\etablx{\end{table*}}
\def\btab{\begin{tabular}} 
\def\btabu{\begin{tabular}}             \def\etabu{\end{tabular}}
\def\btabx{\begin{tabular*}}            \def\etabx{\end{tabular*}}
\def\bbib{\begin{thebibliography}{999}} \def\ebib{\end{thebibliography}}
\def\bver{\begin{verbatim}}             \def\ever{\end{verbatim}}
\def\bca{\begin{cases}}                          \def\eca{\end{cases}}

\def\d#1{\,{\rm d}#1} 
\def\cste{\mbox{cste}} 
\def\pasr{\mbox{pas}}
\def\plg{polygone }
\def\as{angle solide }
\def\plh{poly\`edre }
\def\DP{d\'eriv\'ee partielle }
\def\DPs{d\'eriv\'ees partielles }
\def\ttr{t\'etra\`edre }
\def\mod{\text{mod}}
\def\ku{1)}
\def\kuu{(\ku}
\def\kd{2-a)}
\def\kdd{(\kd}
\def\kt{2-b)}
\def\ktt{(\kt}
\def\pxta{p\left( \xb(1),\ldots,\xb(T) | \Ab \right)}
\def\pxas{p\left( \xb_{1..T} | \Ab, \sb_{1..T} \right)}
\def\pasx{p\left( \Ab,\sb_{1..T} | \xb_{1..T} \right)}
\def\pax{p\left( \Ab | \xb_{1..T} \right)}
\def\psx{p\left( \sb_{1..T} | \xb_{1..T} \right)}
\def\sbh{\wh{\sb}}
\def\abh{\wh{\ab}}
\def\Sbh{\wh{\Sb}}
\def\Abh{\wh{\Ab}}
\def\Bbh{\wh{\Bb}}
\def\zbh{\wh{\zb}}
\def\xbh{\wh{\xb}}
\def\hsp{\hspace*{.5cm}}
\def\hspu{\hspace*{.1cm}}
\def\hspd{\hspace*{.2cm}}
\def\hspt{\hspace*{.3cm}}
\def\hspq{\hspace*{.4cm}}
\def\hspn{\hspace*{-.6cm}}
\def\hspnd{\hspace*{-1.2cm}}
\def\vspq{\vspace*{.4cm}}
\def\vspd{\vspace*{.2cm}}
\def\vspt{\vspace*{.3cm}}
\def\vspu{\vspace*{.1cm}}
\def\vspuu{\vspace*{.05cm}}
\def\vspn{\vspace*{-.1cm}}
\def\vspnd{\vspace*{-.2cm}}
\def\vspnq{\vspace*{-.4cm}}

\bdoc
\begin{abstract}
In this contribution, we present new algorithms to source separation for the case of  noisy instantaneous linear mixture, within the Bayesian statistical framework. The source distribution prior is modeled by a mixture of Gaussians
\cite{Moulines97} and the mixing matrix elements distributions by a Gaussian  \cite{Djafari99a}. We model the mixture of Gaussians hierarchically by  mean of hidden variables representing the labels of the mixture. Then, we consider the joint a posteriori distribution of sources, mixing matrix elements, labels of the mixture and other parameters of the mixture with appropriate prior probability laws to eliminate degeneracy of the likelihood function of variance parameters and we propose two iterative algorithms to estimate jointly sources,  mixing matrix and hyperparameters: Joint MAP (Maximum \apost) algorithm and penalized EM algorithm. The illustrative example is taken in \cite{Macchi99} to compare with other algorithms proposed in literature.
\end{abstract}

\maketitle

\section{Problem description}
We consider a linear instantaneous mixture of $n$ sources. Observations could be corrupted by an additive noise. This noise may represent measurement errors or model incertainty:

\beq\label{model}
\xb(t)=\Ab\sb(t)+\epsilon(t),\qquad t=1,..,T
\eeq

\noindent where $\xb(t)$ is the  ($m \times 1$) measurement vector,  $\sb(t)$ is the ($n \times 1$) source vector which components have to be separated, $\Ab$ is the mixing matrix of dimension ($m \times n$) and  $\epsilon(t)$ represents noise affecting the measurements. We assume that the ($m \times T$) noise matrix $\epsilon(t)$ is statistically independant of sources, centered, white and Gaussian with known variance $\sigma_{\epsilon}^{2}\,\Ib$. We note $\sb_{1..T}$ the matrix $n \times T$ of sources and $\xb_{1..T}$ the matrix $m \times T$ of data.
\\
\\
Source separation problem consists of two sub-problems: Sources restoration and mixing matrix identification. Therefore, three directions can be followed:
\begin{enumerate}
\item {\em Supervised learning}: Identify $\Ab$ knowing a training sequence of sources $\sb$, then use it to reconstruct the sources.
\item {\em Unsupervised learning}: Identify $\Ab$ directly from a part or the whole observations and then use it to recover $\sb$.
\item {\em Unsupervised joint estimation}: Estimate jointly $\sb$ and $\Ab$
\end{enumerate}
In the following, we investigate the third direction. This choice is motivated by  practical cases where sources and mixing matrix are unknown.


This paper is organised as follows: We begin in section II by proposing a Bayesian approach to source separation. We set up the notations, present the prior laws of the sources and the mixing matrix elements and present the joint MAP estimation algorithm assuming known hyperparameters. We introduce, in section III, a hierarchical modelisation of the sources by  mean of  hidden variables representing the labels of the mixture of Gaussians in the prior modeling and present a version of JMAP using the estimation of these hidden variables (classification) as an intermediate step. In both algorithms, we assumed known the hyperparameters which is not realistic in applications. That is why, in section IV, we present an original method for the estimation of  hyperparameters which takes advantages of using this hierarchical modeling. Finally, since EM algorithm has been used extensively in source separation \cite{Bermond00}, we considered this  algorithm and  propose, in section V, a penalized version of the EM algorithm for source separation. This penalization of the likelihood function is necessary to eliminate its degeneracy when some variances of Gaussian mixture approche zero \cite{Ridolfi99}. Each section  is supported by one typical simulation result and partial conclusion. At the end, we compare the two last algorithms.

\section{Bayesian approach to source separation}
Given the observations $\xb_{1..T}$, the joint \apost distribution of unknown variables $\sb_{1..T}$ and $\Ab$ is:
\beq\label{distribution}
\pasx \propto \pxas\,p(\Ab)\,p(\sb_{1..T})
\eeq
\noindent where $p(\Ab)$ and $p(\sb_{1..T})$ are the prior distributions through which we modelise our \aprio information about sources $\sb$ and mixing matrix $\Ab$. $\pxas$ is the joint likelihood distribution. We have, now, three directions:
\begin{enumerate}
\item First, integrate (\ref{distribution}) with respect to $\sb_{1..T}$ to obtain the marginal in $\Ab$ and then estimate  $\Ab$ by:
\beq \label{JMAP2}
\Abh=\argmax_{\Ab}{\{J(\Ab)=\ln \,\pax \}}
\eeq
\item Second, integrate (\ref{distribution})  with respect to $\Ab$ to obtain the marginal in $\sb_{1..T}$ and then estimate $\sb_{1..T}$ by:
\beq \label{JMAP3}
\sbh_{1..T}=\argmax_{\sb_{1..T}}{\{J(\sb_{1..T})=\ln \,\psx \}}
\eeq
\item Third, estimate jointly $\sb_{1..T}$ and  $\Ab$:
\beq \label{JMAP1}
(\Abh,\sbh_{1..T})=\argmax_{(\Ab,\sb_{1..T})}{\{J(\Ab,\sb_{1..T})=\ln \,\pasx\}}
\eeq
\end{enumerate}   

\subsection{Choice of \aprio distributions}
The \aprio distribution  reflects our knowledge concerning the parameter to be estimated. Therefore, it must be neither very specific to a particular problem nor too general (uniform) and non informative. A parametric model for these distributions seems to fit this goal: Its stucture expresses the particularity of the problem and  its parameters allow a certain flexibility.

\noindent{\bf{Sources \aprio:}} For sources $\sb$, we choose a mixture of Gaussians \cite{Moulines97}: 
\beq
p(s_{j})=\sum_{i=1}^{q_{j}}\alpha_{ji}\Nc(m_{ji}, \sigma_{ji}^{2}),\qquad j=1..n
\eeq
Hyperparameters $q_{j}$ are supposed to be known. 

\noindent This choice was motivated by the following points:
\bit
\item It represents a  general class of distributions and is convenient in many digital communications and image processing applications.
\item For a Gaussian likelihood $p\left( \xb_{1..T} | \sb_{1..T}, \Ab \right)$ (considered as a function of $\sb_{1..T}$), the \apost law remains in the same class (conjugate prior). We then have only to update the parameters of the mixture with the data.
\eit
\noindent {\bf{Mixing matrix \aprio:}} To account for some model uncertainty, we assign a Gaussian prior law to each element of the mixing matrix $\Ab$: 
\beq
p(\Ab_{ij})=\Nc(\Mb_{ji}, \sigma_{a,ij}^{2})
\eeq
which can  be interpreted as knowing  every element ($\Mb_{ji}$) with some uncertainty ($\sigma_{a,ij}^{2}$). We underline here the advantage of estimating  the mixing matrix $\Ab$ and not a separating matrix $\Bb$ (inverse of $\Ab$) which is the case of almost all the existing methods for source separation (see for example \cite{Cardoso96}). This approach has at least two advantages: ({\bf{i}}) $\Ab$ does not need to be invertible ($n \neq m$), ({\bf{ii}}) naturally, we have some \aprio information on the mixing matrix not on its inverse which may not exist. 
\vspd

\subsection{JMAP algorithm}
We propose an alternating iterative algorithm to estimate jointly $\sb_{1..T}$ and $\Ab$ by extremizing the log-posterior distribution:
\beq 
\left\{\barr{ccc}
\sbh_{1..T}^{(k)} &=& \argmax_{\sb_{1..T}}{ \ln\,p\left( \Abh^{(k-1)},\sb_{1..T} | \xb_{1..T} \right) }
\\ 
 \Abh^{(k)} &=& \argmax_{\Ab}{ \ln\,p\left( \Ab,\sbh_{1..T}^{(k)} | \xb_{1..T} \right) }
\earr\right.
\eeq
In the following, we suppose that sources are white and spatially independant. This assumption is not necessary in our approach but we start from here to be able to compare later with  other classical methods in which this hypothesis is fundamental.

\noindent With this hypothesis, in step $(k+1)$, the criterion to optimize with respect to $\sb_{1..T}$ is:
\beq
J(\sb_{1..T})=\sum_{t=1}^{T} \left[ \ln\,p\left( \xb(t) | \Abh^{(k)},\sb(t) \right) +\sum_{j=1}^{n}\ln\,p_j\left( \sb_j(t) \right) \right]
\eeq
Therefore, the optimisation is done independantly at each time $t$:
\beq
\sbh(t)^{(k+1)} = \argmax_{\sb(t)}{ \{ \ln\,p\left( \xb(t) | \Abh^{(k)},\sb \right) + \sum_{j=1}^{n} \ln\,p_{j} \left( \sb_{j}(t) \right) \}}
\eeq
The \apost distribution of $\sb$ is a mixture of $\prod_{j=1}^{n}q_j$ Gaussians. This leads to a high computational cost. To obtain a more reasonable algorithm, we propose an iterative scalar algorithm. The first step consists in estimating each source component knowing the other components estimated in the previous iteration:
\beq
\sbh_{j}(t)^{(k+1)} =\argmax_{\sb_{j}(t)}{ \{\ln\,p\left( \sb_{j}(t) | \xb(t),\Abh^{(k)},\sbh_{l \neq  j}(t)^{(k)} \right) \}}    
\eeq  
The \apost distribution of $\sb_j$ is a mixture of $q_j$ Gaussians: $\sum_{z=1}^{q_{j}}\alpha_{jz}^{'}\Nc(m_{jz}^{'}, {\sigma_{jz}^{'}}^{2})$, with:
\beq\left\{\barr{c}\label{zapost}
\displaystyle{m_{jz}^{'}=\frac{\sigma_j^{2}m_{jz}+\sigma_{jz}^{2}m_{j}}{\sigma_j^{2}+\sigma_{jz}^{2}}}\\~\\
\displaystyle{{\sigma_{jz}^{'}}^{2}=\frac{\sigma_j^{2}\,\sigma_{jz}^{2}}{\sigma_j^{2}+\sigma_{jz}^{2}}}\\~\\
\displaystyle{\alpha_{jz}^{'}= \alpha_{jz}\,\sqrt{\frac{1}{\sigma_{jz}^{2}+\sigma_j^{2}}}\Exp{\frac{-1}{2}\frac{1}{\sigma_{jz}^{2}+\sigma_j^{2}}(m_j-m_{jz})^2}}
\earr\right.
\eeq
where
\beq\label{treize}
\left\{\barr{c}
\displaystyle{\sigma_j^2=\frac{\sigma_{\epsilon}^2}{\sum_{i=1}^{m}\Ab_{ij}^2}}
\\~\\
\displaystyle{m_j = \frac{\sum_{i=1}^n \Ab_{ij}\,(\xb_i-\xbh_i)}{\sum_{i=1}^{m}\Ab_{ij}^2}}
\\~\\
\displaystyle{\xbh_i =\sum_{l \neq j}\Ab_{il}\,\sb_l}
\earr\right.
\eeq
\\

If the means $m_{jz}^{'}$ aren't close to each other, we are in the case of a multi-modal distribution. The algorithm to estimate $\sb_j$ is to first compute $\xbh_i$, $i=1, \dots, m$, $m_j$ and $\sigma_j^2$ by (\ref{treize}) and then $\alpha_{jz}^{'}$, ${\sigma_{jz}^{'}}^{2}$ and $m_{jz}^{'}$ by (\ref{zapost}), and select the $m_{jz}^{'}$ for which the ratio $\frac{\alpha_{jz}^{'}}{\sigma_{jz}^{'}}$ is the greatest one.

After a full update of all sources $\sb_{1..T}$, the estimate of $\Ab$ is obtained by optimizing:
\beq 
\barr{lcl} 
J(\Ab)=
\sum_{t=1}^{T}\ln\,p\left( \xb(t) | \Ab,\sbh^{k+1}(t) \right) +\ln\,p\left( \Ab(t) \right) + cte
\earr
\eeq 
which is quadratic in elements of $\Ab$. The  gradient has then a simple expression:
\beq
\frac{\partial{J(\Ab)}}{\partial{\Ab_{i,j}}}=\sum_{t=1}^{T}\frac{1}{\sigma_{\epsilon}^2}\sbh_{j}^{k+1}(t) \left( \xb_{i}(t)-\left[ \Ab\,\sbh^{k+1}(t) \right]_{i}\right)-\frac{1}{\sigma_{a;i,j}^2} \left( \Ab_{i,j}-\Mb_{i,j} \right)
\eeq
Cancelling the gradient to zero and defining $\Lambdab_{i,j}=\frac{\sigma_{\epsilon}^{2}}{\sigma_{a;i,j}^2}$, we obtain the following relation:
\beq\label{matrice}
\left[\sum_{t=1}^{T} \left( \xb(t)-\Ab \sbh^{k+1}(t) \right) \sbh^{k+1}(t)^{T}\right]_{i,j}-\Lambdab_{i,j}\left( \Ab_{i,j}-\Mb_{i,j} \right)=0
\eeq
We define the operator {\bf{Vect}} transforming a matrix to a vector by the concatenation of the transposed rows. Operator {\bf{Mat}} is the inverse of {\bf{Vect}}. Applying operator {\bf{Vect}} to relation (\ref{matrice}), we obtain the following expression:
\beq
Vect\left( \xb_{1..T}(\sbh_{1..T}^{k+1})^{T}\right)+\mu Vect(M)=\left( \mu+S^{*} \right) Vect\Ab
\eeq
\noindent where $\mu$ is a diagonal matrix $(nm \times nm)$ which diagonal vector is $Vect((\Lambda_{i,j})_{i=1..m,j=1..n})$ and $\Sb^{*}$ the matrix $(nm \times nm)$ with block diagonals  $\sbh_{1..T}\sbh_{1..T}^{T}$ estimated at  iteration $(k+1)$. We have finally the explicit estimation of $\Ab$:
\beq
\Abh^{k+1}=Mat\left(\left[ \mu+ S^{*} \right]^{-1}\left[ \mu Vect(\Mb)+Vect\left( \xb_{1..T}(\sbh_{1..T}^{k+1})^{T}\right) \right] \right)
\eeq

To show the faisability of this algorithm, we consider in the following a telecommunication example. For this, we simulated synthetic data with sources described by a mixture of $4$ Gaussians centered at $-3$, $-1$, $1$ and $3$, with the same variance $0.01$ and weighted by 0.3, 0.1, 0.4 and 0.2. The unknown mixing matrix is $\Ab=\left( \barr{cc}1&-0.6\\0.6&1 \earr \right)$. We fixed the \aprio parameters of $\Ab$ to: $\Mb=\left( \barr{cc}1&0\\0&1 \earr \right)$ and $\Lambdab=\left( \barr{cc}150&0.009\\0.009&150 \earr \right)$, meaning that we are nearly sure of diagonal values but we are very uncertain about the other elements of $\Ab$. Noise of variance  $\sigma_{\epsilon}^2=1$ was added to the data. The  figure $1$ illustrates the ability of the algorithm to perform the separation. However, we note that estimated sources are very centered arround the means. This is because  we fixed very low values for the \aprio variances of Gaussian mixture. Thus, the algorithm is sensitive to the \aprio parameters and  exploitation of data is useful. We will see in section IV how to deal with this issue.

\bigskip
\noindent\btabu{@{}c@{}}
\includegraphics[width=\textwidth,height=45mm]{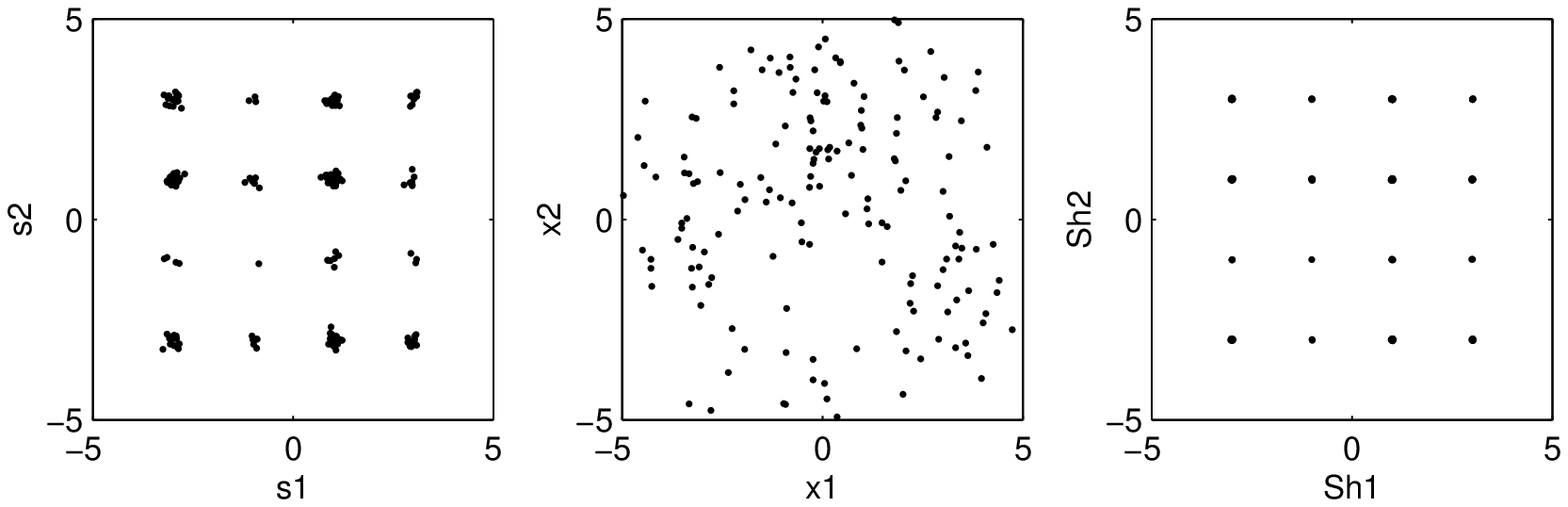}\\
\hspace*{0.5cm}(a)\hspace*{4.5cm}(b)\hspace*{4.5cm}(c)\\~\\
Figure 1- Results of separation with QAM-$16$ (Quadratic Amplitude Modulation) \\
using JMAP algorithm: (a) phase space distribution of sources,\\
(b) mixed signals, and (c) separated sources
\etabu
\\
\\
Now, we are going to re-examine closely the expression for the  \apost distribution of sources. It's a multi-modal distribution if the Gaussian means aren't too close. The maximum of this distribution doesn't correspond, in general, to the maximum of the most probable Gaussian. So, we intend to estimate first, at each time $t$, the \aprio Gaussian law according to which the source $\sb(t)$ is generated (classification) and then estimate $\sb(t)$ as the mean of the \apost Gaussian.  This leads us to the introduction of hidden variables and hierarchical modelization. 

\section{Hidden variables}
The \aprio distribution of the component $s_j$ is  $p(s_{j})=\sum_{i=1}^{q_{j}}\alpha_{ji}\Nc(m_{ji}, \sigma_{ji}^{2})$. We consider now the hidden variable $z_j$ taking its values in the discrete set $\Zc_j=\left( 1, \dots, q_j\right)$  so each source can belong to one of the $q_j$ sources, with $\alpha_{ji}=p\left( z_j=i \right)$. Given $z_j=i$, $\sb_j$  is normal $\Nc(m_{ji},\sigma_{ji}^{2})$. We can extend this notion to vectorial case by considering the vector $\zb=[z_1,\dots,z_{n}]$ taking its values in the set $\Zc=\Pi_{j=1}^{n}\Zc_{j}$. The $\sb$ distribution given $\zb$ is a normal law $p(\sb | \zb)=\Nc(\mb_z,\bf{\Gamma}_z)$ with:
\beq 
\mb_z=[m_{1z_1}, m_{2z_2},\dots, m_{nz_n}]
\eeq
\beq
\bf{\Gamma}_{z}=diag(\sigma_{1z_1}^{2}, \sigma_{2z_2}^{2},\dots, \sigma_{nz_n}^{2})
\eeq
The marginal \aprio law of $\sb$ is the mixture of $\Pi_{j=1}^{n}q_j$ Gaussians:
\beq
p(\sb)=\sum_{\zb \in \Zc} p(\zb)p(\sb | \zb)
\eeq
We can re-interpret this mixture by considering it as a discrete set of couples $\left( \Nc_z, p(\zb) \right)$ (see Figure $2$). Sources which belong to this class of distributions are generated as follows: First, generate the hidden variable $\zb \in \Zc$ according $p(\zb)$ and then, given this $\zb$, generate  $\sb$ according to $\Nc_z$. This model can be extended to include continuous values of $\zb$ (also continuous distribution $f(\zb)$) and then to take account of infinity of distributions in only one class  (see Figure $2$).
\\
\bcc 
\fbox{\setlength{\unitlength}{0.00023333in}
\begingroup\makeatletter\ifx\SetFigFont\undefined%
\gdef\SetFigFont#1#2#3#4#5{%
  \reset@font\fontsize{#1}{#2pt}%
  \fontfamily{#3}\fontseries{#4}\fontshape{#5}%
  \selectfont}%
\fi\endgroup%
{\renewcommand{\dashlinestretch}{30}
\begin{picture}(10821,7839)(0,-10)
\thicklines
\path(10434,5037)(10359,4812)
\path(7756,605)(7681,380)
\thinlines
\path(1209,6162)(1584,6612)
\blacken\path(1530.225,6500.608)(1584.000,6612.000)(1484.131,6539.019)(1530.225,6547.469)(1530.225,6500.608)
\path(2484,6462)(3084,6387)
\blacken\path(2961.206,6372.116)(3084.000,6387.000)(2968.648,6431.653)(3000.649,6397.419)(2961.206,6372.116)
\path(1663,5246)(2338,5396)
\blacken\path(2227.365,5340.683)(2338.000,5396.000)(2214.350,5399.254)(2256.000,5377.778)(2227.365,5340.683)
\blacken\path(2933.845,5838.520)(2859.000,5937.000)(2878.696,5814.885)(2892.089,5859.792)(2933.845,5838.520)
\path(2859,5937)(3084,5412)
\path(5409,5262)(10809,5262)
\blacken\thicklines
\path(10689.000,5209.500)(10809.000,5262.000)(10689.000,5314.500)(10725.000,5262.000)(10689.000,5209.500)
\thinlines
\blacken\thicklines
\path(6888.889,7693.074)(6834.000,7812.000)(6783.910,7690.974)(6835.680,7728.017)(6888.889,7693.074)
\path(6834,7812)(6909,4062)(6834,4062)
\thinlines
\blacken\thicklines
\path(7471.500,6267.000)(7434.000,6387.000)(7396.500,6267.000)(7434.000,6303.000)(7471.500,6267.000)
\path(7434,6387)(7434,5187)
\thinlines
\blacken\thicklines
\path(8071.500,5817.000)(8034.000,5937.000)(7996.500,5817.000)(8034.000,5853.000)(8071.500,5817.000)
\path(8034,5937)(8034,5262)
\thinlines
\path(5184,4062)(5184,2787)
\blacken\thicklines
\path(5124.000,2907.000)(5184.000,2787.000)(5244.000,2907.000)(5184.000,2871.000)(5124.000,2907.000)
\thinlines
\blacken\thicklines
\path(5206.500,2367.000)(5184.000,2487.000)(5161.500,2367.000)(5184.000,2403.000)(5206.500,2367.000)
\path(5184,2487)(5184,12)
\thinlines
\path(4434,837)(8184,837)
\blacken\thicklines
\path(8064.000,814.500)(8184.000,837.000)(8064.000,859.500)(8100.000,837.000)(8064.000,814.500)
\thinlines
\blacken\thicklines
\path(8589.000,5667.000)(8559.000,5787.000)(8529.000,5667.000)(8559.000,5703.000)(8589.000,5667.000)
\path(8559,5787)(8559,5262)
\thinlines
\path(1659,7212)(1657,7211)(1653,7209)
	(1646,7206)(1635,7201)(1619,7193)
	(1598,7183)(1571,7170)(1539,7154)
	(1502,7136)(1461,7115)(1415,7093)
	(1365,7068)(1314,7042)(1260,7015)
	(1205,6987)(1149,6958)(1094,6929)
	(1040,6901)(987,6872)(935,6844)
	(886,6817)(838,6790)(793,6763)
	(750,6738)(709,6713)(670,6689)
	(634,6665)(599,6642)(567,6619)
	(536,6596)(507,6573)(479,6551)
	(453,6529)(428,6506)(404,6483)
	(381,6460)(359,6437)(336,6412)
	(315,6386)(294,6360)(273,6334)
	(254,6306)(234,6278)(216,6249)
	(198,6220)(181,6190)(165,6159)
	(149,6127)(134,6095)(120,6063)
	(106,6029)(94,5996)(82,5962)
	(71,5927)(61,5893)(52,5858)
	(44,5823)(37,5788)(30,5754)
	(25,5720)(20,5686)(17,5652)
	(14,5619)(13,5586)(12,5553)
	(12,5522)(13,5490)(14,5459)
	(17,5429)(20,5399)(24,5370)
	(29,5341)(34,5312)(40,5282)
	(48,5252)(56,5222)(65,5191)
	(75,5161)(86,5131)(98,5101)
	(111,5070)(126,5039)(141,5009)
	(157,4978)(174,4947)(193,4916)
	(212,4886)(232,4855)(253,4825)
	(276,4795)(298,4766)(322,4737)
	(347,4709)(372,4681)(397,4654)
	(423,4627)(450,4602)(477,4577)
	(504,4553)(532,4529)(560,4507)
	(588,4485)(616,4464)(645,4444)
	(674,4424)(704,4405)(734,4387)
	(763,4370)(792,4353)(823,4337)
	(854,4321)(886,4306)(918,4290)
	(952,4275)(987,4261)(1022,4246)
	(1059,4232)(1096,4219)(1134,4206)
	(1174,4193)(1214,4181)(1254,4170)
	(1296,4159)(1337,4149)(1380,4139)
	(1422,4131)(1465,4122)(1509,4115)
	(1552,4108)(1595,4103)(1638,4098)
	(1681,4094)(1724,4090)(1766,4088)
	(1808,4086)(1850,4085)(1891,4085)
	(1932,4085)(1973,4086)(2014,4089)
	(2054,4091)(2094,4095)(2134,4099)
	(2172,4104)(2210,4110)(2249,4116)
	(2288,4123)(2327,4131)(2367,4140)
	(2407,4149)(2447,4160)(2488,4171)
	(2530,4183)(2571,4196)(2613,4210)
	(2656,4224)(2698,4240)(2741,4256)
	(2783,4273)(2826,4291)(2868,4309)
	(2910,4329)(2951,4348)(2992,4369)
	(3033,4390)(3073,4412)(3112,4434)
	(3150,4456)(3188,4479)(3224,4502)
	(3260,4526)(3294,4550)(3328,4574)
	(3360,4598)(3391,4623)(3422,4647)
	(3451,4672)(3479,4697)(3507,4723)
	(3533,4748)(3559,4774)(3585,4802)
	(3611,4831)(3636,4860)(3660,4890)
	(3684,4920)(3707,4952)(3730,4984)
	(3752,5017)(3773,5051)(3794,5085)
	(3814,5121)(3834,5157)(3853,5193)
	(3871,5231)(3889,5269)(3906,5307)
	(3922,5346)(3937,5385)(3952,5424)
	(3965,5463)(3978,5503)(3990,5542)
	(4001,5581)(4011,5620)(4021,5658)
	(4029,5696)(4037,5733)(4043,5771)
	(4049,5807)(4055,5843)(4059,5879)
	(4063,5914)(4066,5948)(4068,5982)
	(4070,6016)(4072,6049)(4072,6087)
	(4072,6124)(4071,6162)(4070,6200)
	(4068,6237)(4064,6275)(4060,6314)
	(4055,6352)(4050,6390)(4043,6429)
	(4035,6467)(4027,6506)(4017,6544)
	(4007,6582)(3996,6619)(3983,6656)
	(3970,6692)(3957,6727)(3942,6762)
	(3927,6795)(3910,6827)(3894,6858)
	(3876,6888)(3859,6917)(3840,6944)
	(3821,6970)(3802,6994)(3782,7018)
	(3761,7040)(3740,7061)(3719,7081)
	(3697,7099)(3674,7117)(3650,7134)
	(3625,7150)(3600,7166)(3573,7180)
	(3546,7194)(3517,7207)(3487,7219)
	(3457,7231)(3425,7242)(3392,7252)
	(3359,7262)(3325,7271)(3290,7279)
	(3254,7287)(3218,7293)(3182,7299)
	(3146,7305)(3109,7310)(3072,7314)
	(3036,7317)(3000,7320)(2964,7322)
	(2928,7324)(2893,7325)(2858,7326)
	(2824,7327)(2790,7327)(2757,7327)
	(2724,7326)(2691,7325)(2659,7324)
	(2625,7323)(2590,7322)(2556,7320)
	(2521,7318)(2487,7316)(2452,7313)
	(2417,7311)(2381,7308)(2346,7305)
	(2310,7301)(2275,7298)(2240,7294)
	(2205,7290)(2171,7286)(2137,7282)
	(2104,7277)(2071,7273)(2040,7268)
	(2010,7264)(1980,7259)(1952,7254)
	(1926,7249)(1900,7245)(1876,7240)
	(1854,7235)(1832,7230)(1812,7226)
	(1793,7221)(1776,7217)(1759,7212)
	(1731,7203)(1706,7195)(1685,7185)
	(1667,7176)(1651,7165)(1637,7154)
	(1625,7141)(1614,7127)(1605,7113)
	(1598,7099)(1592,7087)(1588,7076)
	(1586,7069)(1585,7064)(1584,7062)
\path(4734,1437)(4736,1438)(4740,1440)
	(4746,1444)(4757,1450)(4772,1459)
	(4792,1470)(4816,1484)(4844,1499)
	(4875,1517)(4910,1536)(4947,1556)
	(4985,1577)(5025,1598)(5064,1618)
	(5103,1638)(5140,1656)(5177,1673)
	(5211,1688)(5244,1701)(5275,1712)
	(5305,1721)(5333,1728)(5359,1733)
	(5384,1736)(5409,1737)(5435,1736)
	(5462,1733)(5488,1728)(5514,1721)
	(5540,1712)(5566,1702)(5592,1690)
	(5619,1677)(5645,1662)(5671,1647)
	(5697,1631)(5723,1615)(5749,1599)
	(5775,1582)(5800,1566)(5825,1549)
	(5850,1534)(5875,1518)(5899,1503)
	(5922,1489)(5945,1475)(5967,1462)
	(5988,1450)(6009,1437)(6033,1422)
	(6056,1407)(6077,1392)(6098,1376)
	(6118,1360)(6137,1345)(6155,1329)
	(6173,1312)(6190,1296)(6207,1280)
	(6224,1264)(6241,1248)(6258,1232)
	(6275,1217)(6293,1202)(6310,1188)
	(6328,1174)(6347,1161)(6365,1149)
	(6384,1137)(6403,1126)(6421,1116)
	(6440,1106)(6458,1097)(6476,1089)
	(6494,1081)(6512,1073)(6529,1066)
	(6547,1058)(6564,1051)(6582,1045)
	(6600,1038)(6618,1031)(6636,1025)
	(6655,1018)(6675,1012)(6695,1005)
	(6715,999)(6737,993)(6759,987)
	(6780,982)(6801,977)(6823,973)
	(6846,969)(6869,965)(6893,961)
	(6917,957)(6942,953)(6967,949)
	(6993,946)(7018,942)(7043,939)
	(7069,936)(7094,933)(7119,929)
	(7144,926)(7168,924)(7192,921)
	(7216,918)(7239,916)(7262,914)
	(7284,912)(7308,910)(7333,909)
	(7358,908)(7385,907)(7413,907)
	(7443,907)(7474,907)(7506,907)
	(7540,908)(7573,908)(7605,909)
	(7635,910)(7662,910)(7685,911)
	(7704,911)(7717,912)(7727,912)
	(7732,912)(7734,912)
\path(4734,1437)(4732,1435)(4727,1432)
	(4719,1425)(4707,1415)(4691,1401)
	(4671,1384)(4647,1364)(4621,1343)
	(4593,1319)(4564,1296)(4536,1273)
	(4508,1250)(4482,1229)(4457,1210)
	(4434,1192)(4414,1176)(4394,1162)
	(4376,1149)(4359,1137)(4340,1125)
	(4322,1114)(4305,1103)(4288,1094)
	(4271,1084)(4254,1076)(4238,1067)
	(4222,1059)(4205,1051)(4189,1044)
	(4173,1036)(4157,1029)(4141,1022)
	(4125,1014)(4109,1007)(4092,1000)
	(4076,993)(4059,987)(4040,980)
	(4020,974)(3999,968)(3976,961)
	(3952,955)(3925,949)(3897,942)
	(3869,936)(3842,930)(3817,924)
	(3796,920)(3780,916)(3768,914)
	(3762,913)(3759,912)
\put(534,5937){\makebox(0,0)[lb]{\smash{{{\SetFigFont{6}{6}{\rmdefault}{\bfdefault}{\updefault}($N_1$,
$p_1$)}}}}}
\put(2259,6687){\makebox(0,0)[lb]{\smash{{{\SetFigFont{6}{6}{\rmdefault}{\bfdefault}{\updefault}($N_2$,
$p_2$)}}}}}
\put(1209,4887){\makebox(0,0)[lb]{\smash{{{\SetFigFont{6}{6}{\rmdefault}{\bfdefault}{\updefault}($N_3$,
$p_3$)}}}}}
\put(2334,3612){\makebox(0,0)[lb]{\smash{{{\SetFigFont{6}{6}{\rmdefault}{\mddefault}{\updefault}$\Fc$}}}}}
\put(7359,4887){\makebox(0,0)[lb]{\smash{{{\SetFigFont{6}{6}{\rmdefault}{\mddefault}{\updefault}$1$}}}}}
\put(7959,4887){\makebox(0,0)[lb]{\smash{{{\SetFigFont{6}{6}{\rmdefault}{\mddefault}{\updefault}$2$}}}}}
\put(8484,4887){\makebox(0,0)[lb]{\smash{{{\SetFigFont{6}{6}{\rmdefault}{\mddefault}{\updefault}$3$}}}}}
\put(10434,4812){\makebox(0,0)[lb]{\smash{{{\SetFigFont{6}{6}{\rmdefault}{\mddefault}{\itdefault}R}}}}}
\put(7134,7587){\makebox(0,0)[lb]{\smash{{{\SetFigFont{6}{6}{\rmdefault}{\mddefault}{\updefault}$p(z)$}}}}}
\put(5334,2262){\makebox(0,0)[lb]{\smash{{{\SetFigFont{6}{6}{\rmdefault}{\mddefault}{\updefault}$p(z)$}}}}}
\put(7734,387){\makebox(0,0)[lb]{\smash{{{\SetFigFont{6}{6}{\rmdefault}{\mddefault}{\itdefault}R}}}}}
\put(3834,3687){\makebox(0,0)[lb]{\smash{{{\SetFigFont{6}{6}{\rmdefault}{\mddefault}{\updefault}generalize}}}}}
\end{picture}
}
}\\~\\
Figure 2- Hierarchical modelization with hidden variables
\ecc

\subsection{\apost distribution of sources}
In the following, we suppose that mixing matrix is known. The joint law of $\sb$, $\zb$ and $\xb$ can be  factorized in two forms: $p(\sb,\zb,\xb)=p(\xb|\sb)p(\sb|\zb)p(\zb)$ or $p(\sb,\zb,\xb)=p(\sb | \xb,\zb)p(\zb|\xb)p(\xb)$. Thus, the marginal \apost law has two forms:
\beq
p(\sb | \xb)=\sum_{\zb \in \Zc}\frac{p(\zb)\,p(\xb|\sb)\,p(\sb|\zb)}{p(\xb)}
\eeq
or
\beq
p(\sb | \xb)=\sum_{\zb \in \Zc}p(\zb|\xb)\,p(\sb | \xb,\zb)
\eeq
We note in the second form that the \apost is in the same class that of the \aprio (same expressions but conditionally to $\xb$). This is due to the fact that mixture of Gaussians is a conjugate prior for Gaussian likelihood. Our strategy of estimation is based on this remark: The  sources are modeled hierarchically, we estimate them hierarchically; we begin by estimating the hidden variable using $p(\zb|\xb)$  and then estimate sources using $p(\sb | \xb,\zb)$ which is Gaussian of mean $\thetab_{xz}$:
\beq\label{moyenne}
\thetab_{xz}=\mb_{z}+\bf{\Gamma}_{z}\Ab^{t}R_z(\xb-\Ab \mb_z)
\eeq
and variance $\Vb_{xz}$:
\beq\label{variance}
\Vb_{xz}=\bf{\Gamma}_z-\bf{\Gamma}_z\Ab^{t}R_z\Ab \bf{\Gamma}_z
\eeq
where,
\beq
\Rb_{z}=(\Ab\bf{\Gamma}_z\Ab^{t}+ \Rb_n)^{-1}
\eeq
and $\Rb_n$ represent the noise covariance.

Now we have to estimate $\zb$ by using $p(\zb | \xb)$ which is  obtained by integrating the joint \apost of $\zb$ and $\sb$ with respect to $\sb$:
\beq
p(\zb | \xb) = \int p(\zb , \sb | \xb) d\sb \,\propto\, p(\zb)\int p(\xb|\sb)\,p(\sb | \zb) d\sb
\eeq

The expression to integrate is Gaussian in $\sb $. The result is immediate:

\beq
p(\zb | \xb) \propto p(\zb)\mid \Gammab_z \mid ^{-\frac{1}{2}}\mid \Vb_{xz} \mid ^{\frac{1}{2}} \Exp{K_{zx}}
\eeq
where:
\beq
\left\{\barr{ccc}
\Kb_{zx} &=& -\frac{1}{2}(\Ab \mb_z-\xb)^{t}\Qb_{xz}(\Ab \mb_z-\xb)
\\
\Qb_{xz} &=& (\Ib-\Rb_z\Ab \bf{\Gamma}_z \Ab^{t})\Rb_{n}^{-1}(\Ib-\Ab \bf{\Gamma}_z\Ab^{t}R_z)+R_z\Ab\bf{\Gamma}_z\Ab^{t}R_z
\earr\right\}
\eeq

\noindent If now we consider the whole observations, the law of $\zb_{1..T}$ is:
\beq
p(\zb_{1..T} | \xb_{1..T})\propto p(\zb_{1..T}) \int p(\xb_{1..T}|\sb_{1..T})\, p(\sb_{1..T}|\zb_{1..T}) \,d\sb_{1..T}
\eeq
Supposing that $\zb(t)$ are \aprio independant, the last relation becomes:
\beq
p(\zb_{1..T} | \xb_{1..T})\propto \Pi_{t=1}^{T}\left\{ p(\zb(t)) \int p(\xb(t)|\sb(t)) \,p(\sb(t)|\zb(t))\, d\sb(t) \right\}
\eeq  
Estimation of $\zb_{1..T}$ is then performed observation by observation:
\beq
\argmax_{\zb_{1..T}}{p(\zb_{1..T} | \xb_{1..T})}=\left( \argmax_{\zb(t)}{p(\zb(t) | \xb(t))} \right)_{t=1..T}
\eeq

\subsection{Hierarchical JMAP algorithm}
Taking into account of this hierarchical model, the JMAP algorithm is implemented in three steps. At iteration $(k)$:
\begin{enumerate}
\item First, estimate the hidden variable $\zbh_{MAP}$ (combinatary estimation) given observations and mixing matrix estimated in the previous iteration:
\beq
\zbh_{MAP}^{(k)}(t)=\argmax_{\zb(t)}{\{p\left(\zb(t)|\xb(t), \Abh^{(k-1)}\right)\}}
\eeq
\item Second, given the estimated $\zbh_{MAP}^{(k)}$, source vector $\sb$ follows Gaussian law \\
$\Nc(\thetab_{x\zbh_{MAP}^{(k)}},\Vb_{x\zbh_{MAP}^{(k)}})$ and then the source estimate is $\thetab_{x\zbh_{MAP}^{(k)}}$.
\item Third, given the estimated sources $\sbh^{k}$, mixing matrix is evaluated as in the algorithm of section II.
\end{enumerate}
We evaluated this algorithm using the same  synthetic data as in  section $2$. Separation was robust as shown in Figure $3$:  

\bigskip
\noindent\btabu{@{}c@{}}
\includegraphics[width=\textwidth,height=45mm]{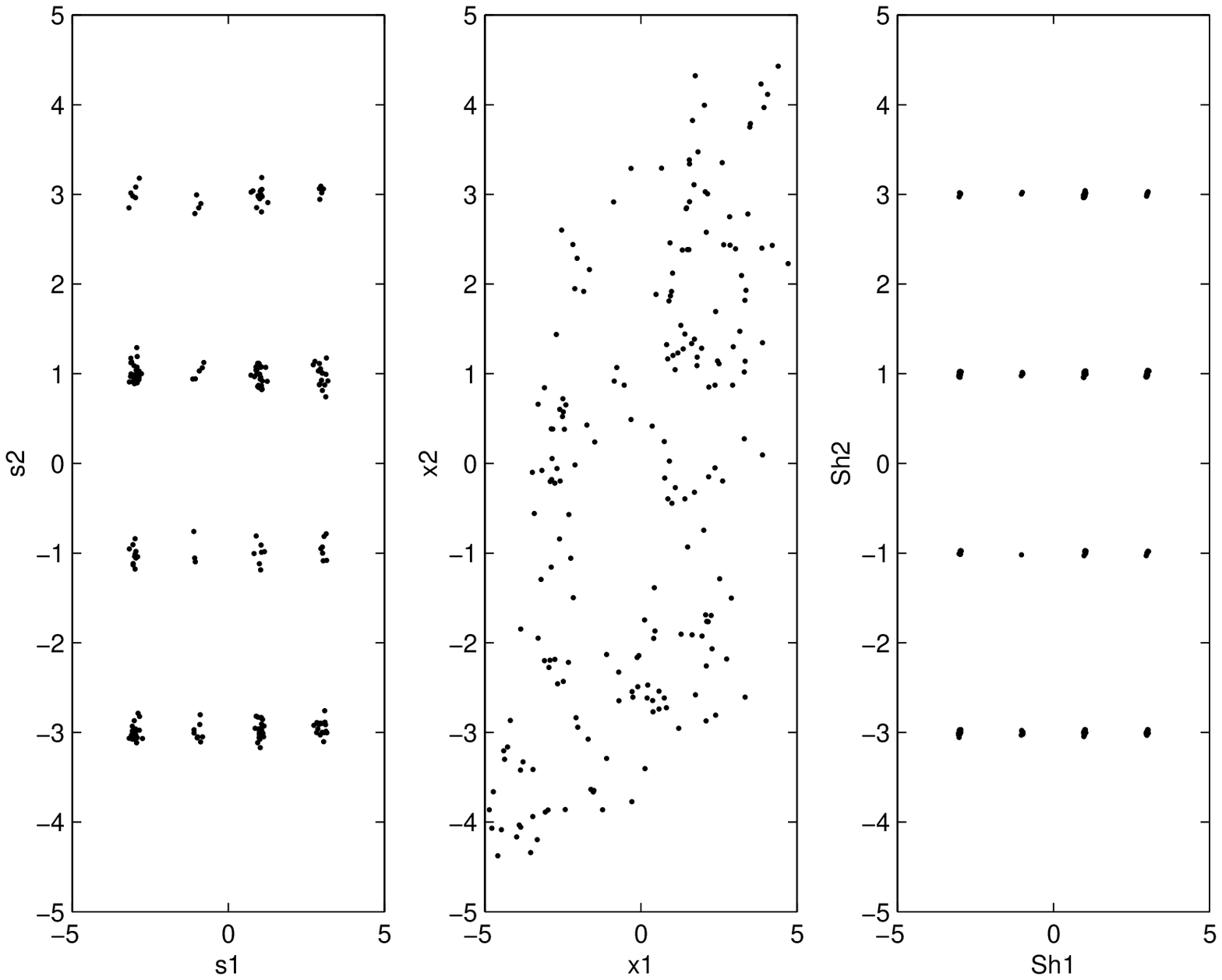}\\
\hspace*{0.5cm}(a)\hspace*{4.5cm}(b)\hspace*{4.5cm}(c)\\~\\
Figure 3- Results of separation with QAM-$16$ \\
using Hierarchical JMAP algorithm: (a) phase space distribution of sources,\\
(b) mixed signals, and (c) separated sources
\etabu 
\\
\\  
The Bayesian approach allows us to express our \aprio information via parametric prior models. However, in general, we may not know the parameters of the \aprio distributions. This is the task of the next section where we estimate the unknown hyperparameters always in a Bayesian framework.   

\section{Hyperparameters estimation}
The hyperparameters considered here are the means and the variances of Gaussian mixture prior of sources: $
\sb_j \sim \sum_{z=1}^{q_j}\Pi_{jz} \Nc \left( m_{jz}, \frac{1}{\psi_{jz}} \right)$, $j=1,\dots, n$. 
We develop, in the following, a novel method to extract the hyperparameters from the observations $\xb_{1..T}$. The main idea is:  
conditioned on the hidden variables $(z_{j})_{1..T}=[z_j(1),\dots, z_j(T)]$,  hyperparameters $m_{jz}$ and $\psi_{jz}$ for $z \in \Zc_j=(1,\dots,q_{j})$ are means and variances of a Gaussian distribution. Thus, given the vector $(z_{j})_{1..T}=[z_j(1),\dots, z_j(T)]$, we can perform a partition of the set $\Tc=[1,\dots, T]$ into sub-sets $\Tc_z$ as:
\beq
\Tc_z=\left \{\, t \,| \,z_j(t) = z \right \} \, ,\,  \, z \in \Zc_j
\eeq
This is the classification step.

Suppose now  that mixing matrix $\Ab$ and components $\sb_{l \neq j}$ are fixed and we are interested in the estimation of $m_{jz}$ and $\psi_{jz}$. Let $\theta_{jz}=( m_{jz}\,,\, \psi_{jz} )$.\\

The joint \apost law of $\sb_j$ and $\theta_{jz}$  given $z_j$ at time  $t$ is:
\beq
p( \sb_j,\,\theta_{jz} \,|\, \xb,\,z_j) \propto p(\xb \,|\, \sb)\,p(\sb_j\,|\, \theta_{jz},\,z_j)\,p(\theta_{jz} \,|\,z_j)
\eeq
$p(\sb_j\,|\, \theta_{jz},\,z_j)$ is Gaussian of mean $m_{jz}$ and inverted variance $\psi_{jz}$.\\
$p(\theta_{jz} \,|\,z_j)\,=\,p(\theta_{jz})\,=\,p(m_{jz})\,p(\psi_{jz})$ is hyperparameters \aprio.
The marginal \apost distribution of $\theta_{jz}$ is obtained from previous relation by integration over $\sb_j$:
\beq
p(\theta_{jz} \,|\, \xb,\,z_j) \propto p(\theta_{jz})\int_{\sb_j}p(\xb \,|\, \sb)\,p(\sb_j\,|\, \theta_{jz},\,z_j)\,d\sb_j.
\eeq
The expression inside the integral is proportional to the joint \apost distribution of $(\sb_j\,,\,z_j)$ given $\xb$ and $\theta_{jz}$, thus:
\beq
p(\theta_{jz} \,|\, \xb,\,z_j) \propto p(\theta_{jz})\,p(z_j\,|\,\xb,\,\theta_{jz}) .
\eeq
where $p(z_j\,|\,\xb,\,\theta_{jz})$ is proportional to $\alpha_{jz}^{'}$ as defined in  expression (\ref{zapost}). Noting $\phi_j=1\,/\,\sigma_j^{2}$ and $\psi_{jz}=1\,/\,\sigma_{jz}^{2}$, we have:
\beq\label{vraisembance}
p(\theta_{jz} \,|\, \xb,\,z_j) \propto p(\theta_{jz})\sqrt{\frac{\phi_j\,\psi_{jz}}{\phi_j\,+\,\psi_{jz} }}\,\Exp{-\frac{1}{2}\,\frac{\phi_j\,\psi_{jz}}{\phi_j+\psi_{jz} }\,(m_{jz}-m_j)^2}
\eeq
Note that the likelihood is normal for means $m_{jz}$ and Gamma  for $\lambda_{jz}=\left(\phi_j\psi_{jz}\right)/\left(\phi_j+
\psi_{jz}\right) $. 

\noindent Choosing a uniform \aprio for the means, the estimate of $m_{jz}$ is:
\beq\label{mean}
\wh{m}_{jz}^{MAP} = \frac{\sum_{t \in \Tc_z}m_j(t)}{T_z} 
\eeq
For variances, we can choose ({\bf i}) an inverted Gamma prior $\Gc\left( \alpha, \beta \right)$ after developing the expression for $\lambda_{jz}$ knowing the relative order of $\psi_{jz}$ and $\phi_j$ (to make $\lambda_{jz}$ linear in $\psi_{jz}$) or ({\bf ii}) an a prior which is  Gamma in $\lambda_{jz}$. These choices are motivated by two points: First, it is a proper prior which eliminate degenaracy of some variances at zero (It is shown in \cite{Ridolfi99} that hyperparameter likelihood (noiseless case without mixing) is unbounded causing a variance degeneracy at zero). Second, it is a conjugate prior so estimation expressions remain simple to implement. The estimate of inverted variance (first choice when $\psi_{jz}$ is the same order of $\phi_j$) is:
\beq\label{variance2}
\wh{\psi}_{jz}^{MAP} = \frac{\alpha^{posteriori}-1}{\beta^{posteriori}}
\eeq
with $\alpha^{posteriori}=\alpha+\frac{\Tc_z}{2}$  and $\beta^{posteriori}=\beta+\frac{\sum_{t \in \Tc_z}(m_j(t)-\wh{m}_{jz}^{MAP})^2}{4}$. 
\vspd
\vspq

\subsection{Hierarchical JMAP  including estimation of hyperparameters}
Including the estimation of hyperparameters, the proposed hierarchical JMAP algorithm is composed of five steps:
\begin{enumerate}
\item Estimate hidden variables $(\wh{z}_j)_{1..T}^{MAP}$ by:
\beq
(\wh{z}_j)_{1..T}^{MAP} = 
(\argmax_{z_j}{p(z_j\,|\,\xb(t),\,m_{jz}\,,\,\psi_{jz},\,\Ab,\,\sb_{l \neq j})})_{1..T}   
\eeq
which permits to estimate  partitions:
\beq
\wh{\Tc_z} = \left \{ t \, |\, (\wh{z}_j)^{MAP}(t)=z \right\}
\eeq
This corresponds to the classification step in the previous algorithm
\item Given the estimate of partitions, hyperparameters $\wh{\psi}_{jz}^{MAP}$ and $\wh{m}_{jz}^{MAP}$ are updated according to equations (\ref{mean}) and (\ref{variance}).
The following steps are the same as those in the previous proposed algorithm
\item Re-estimation of hidden variables $(\wh{z}_j)_{1..T}^{MAP}$ given the estimated hyperparameters.
\item Estimation of sources  $(\wh{\sb})_{1..T}^{MAP}$.
\item Estimation of mixing matrix $(\wh{\Ab})^{MAP}$.
\end{enumerate}
\vspq

\subsection{Simulation results}
To be able to compare the results obtained by this algorithm and the Penalized likelihood algorithm developed in the next section with the results obtained by some other classical methods, we generated data according to the example described in \cite{Macchi99}.
 
\noindent{\bf{Data generation}}: $2$-D sources, every component \aprio is mixture of two Gaussians ($\pm 1$), $\psi=100$ for all Gaussians. Original sources are mixed with mixing matrix $\Ab=\left( \barr{cc} 1&-0.6\\0.4&1 \earr \right)$. A noise of variance $\sigma_{\epsilon}^2 = 0.03$ is added ($SNR=15\,dB$). Number of observations is $1000$.

\noindent {\bf{Parameters}}: $\Mb=\left( \barr{cc} 1&0\\0&1 \earr \right)$,  $\Lambdab=\left( \barr{cc} 150&0.009\\0.009&150 \earr \right)$, $\Pi=\left( \barr{cc} 0.5&0.5\\0.5&0.5 \earr \right)$, $\alpha=200$ and $\beta=2$.  

\noindent {\bf{Initial conditions}}: $\Ab^{(0)}=\left(\barr{cc}1 &0 \\0 &1\earr\right)$, $\psi^{(0)}=\left(\barr{cc}1 &1 \\1 &1\earr\right)$, $m^{(0)}=\left(\barr{cc}0 &0 \\0 &0\earr\right)$ and $\sb^{(0)}$ generated according to  $\sb_j^{(0)} \sim \sum_{z=1}^{q_j}\Pi_{jz} \Nc(m_{jz}^{(0)}, \frac{1}{\psi_{jz}^{(0)}})$.

Sources are recovered with negligible mean quadratic error: 
$MEQ(\sb_1)=0.0094$ and $MEQ(\sb_2)=0.0097$. The following figures illustrate separation results:

The non-negative performance index of \cite{Moreau96} is used to chacarterize mixing matrix identification achievement:
\[
ind(S=\Abh^{-1}\,\Ab) = \frac{1}{2} \left[ \sum_i \left( \sum_j\frac{|S_{ij}|^2}{max_{l}|S_{il}|^2}-1 \right)+\sum_j \left( \sum_i\frac{|S_{ij}|^2}{max_{l}|S_{lj}|^2}-1 \right) \right]
\]
Figure $7a$ represents the index evolution through iterations. Note the convergence of JMAP algorithm since iteration $30$ to a satisfactory value of $-45\,dB$. For the same SNR, algorithms PWS, NS \cite{Macchi99} and EASI \cite{Cardoso96} reach a value greater than $-35\,dB$ after $6000$ observations.
Figures $7b$ and $7c$  illustrate the identification of hyperparameters. We note the algorithm convergence to the original values ($-1$ for $m_{11}$ and $100$ for $\psi_{11}$).
In order to validate the idea of data classification before estimating hyperparameters, we can visualize the evolution of classification error (number of data badly classified). Figure $7d$ shows that this error converges to zero at iteration $15$. Then, after this iteration, hyperparameters identification is performed on the true classified data. Estimation of $m_{jz}$ and $\psi_{jz}$ takes into account only data which belong to this class and then it is not corrupted by other data which bring erroneous information on these hyperparameters. 

\bigskip
\bcc
\btabu[b]{@{}cc@{}}
\btabu[b]{@{}c@{}}
$\left\{\barr{@{}l@{}}
s_1(t) \\ ~\\~\\   
s_2(t)  
\earr\right.$  
~\\ ~\\ 
$\left\{\barr{@{}l@{}}
x_1(t) \\ ~\\~\\   
x_2(t)  
\earr\right.$ 
~\\ ~\\
$\left\{\barr{@{}l@{}}
\wh{s}_1(t) \\ ~\\ ~\\   
\wh{s}_2(t)     
\earr\right.$ 
~\\ ~\\
$\left\{\barr{@{}l@{}}
\wh{s}_1(t)-s_1(t) \\ ~\\ ~\\ 
\wh{s}_2(t)-s_2(t)     
\earr\right.$ 
\\ ~\\ 
\etabu
&
\btabu[b]{@{}c@{}}
\includegraphics[width=100mm,height=100mm]{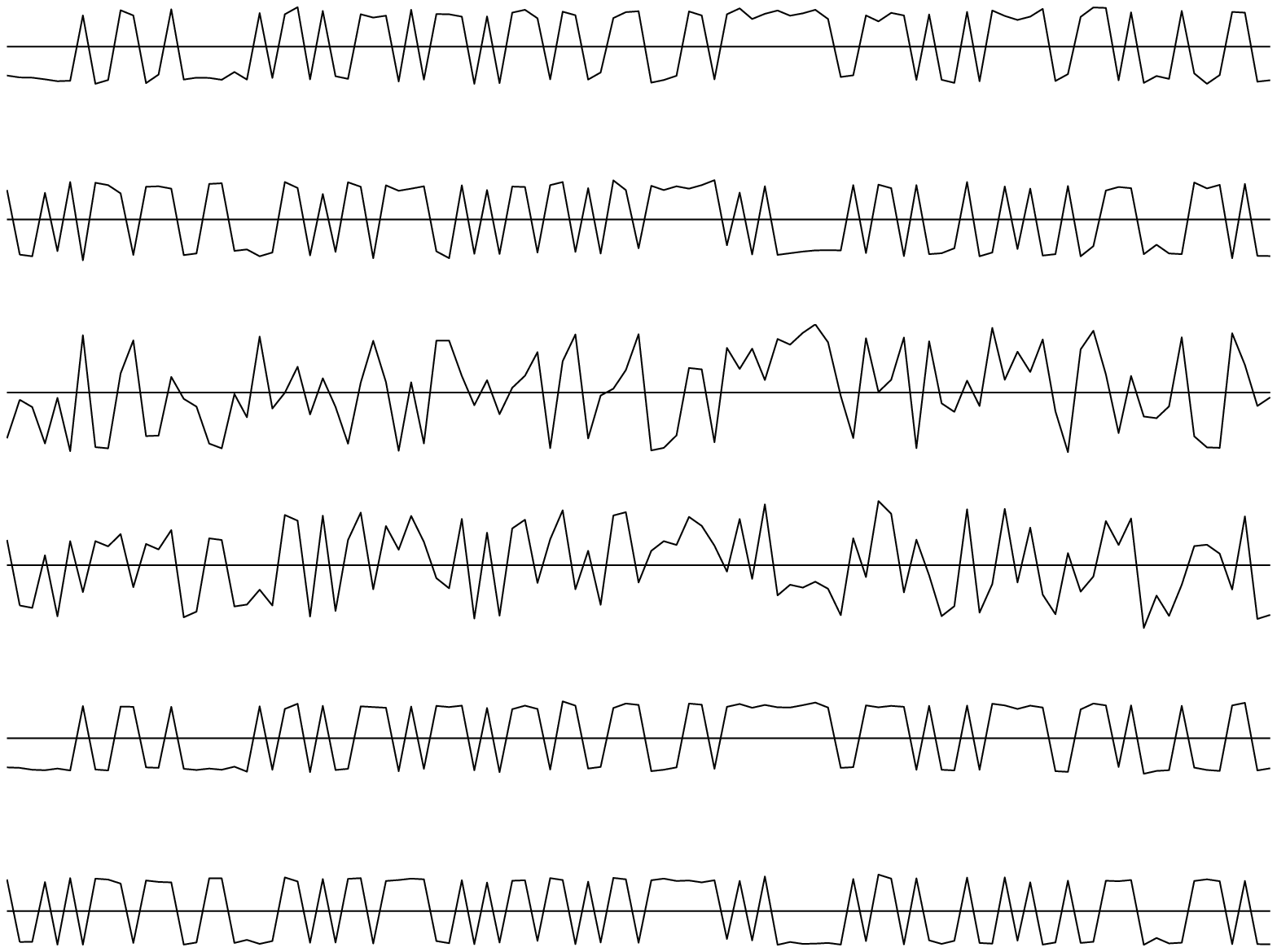}\\
\etabu
\etabu\\~\\
Figure 4- Separation results with $SNR=15\,dB$
\ecc

\bigskip
\noindent\btabu{@{}c@{}}
\includegraphics[width=\textwidth,height=45mm]{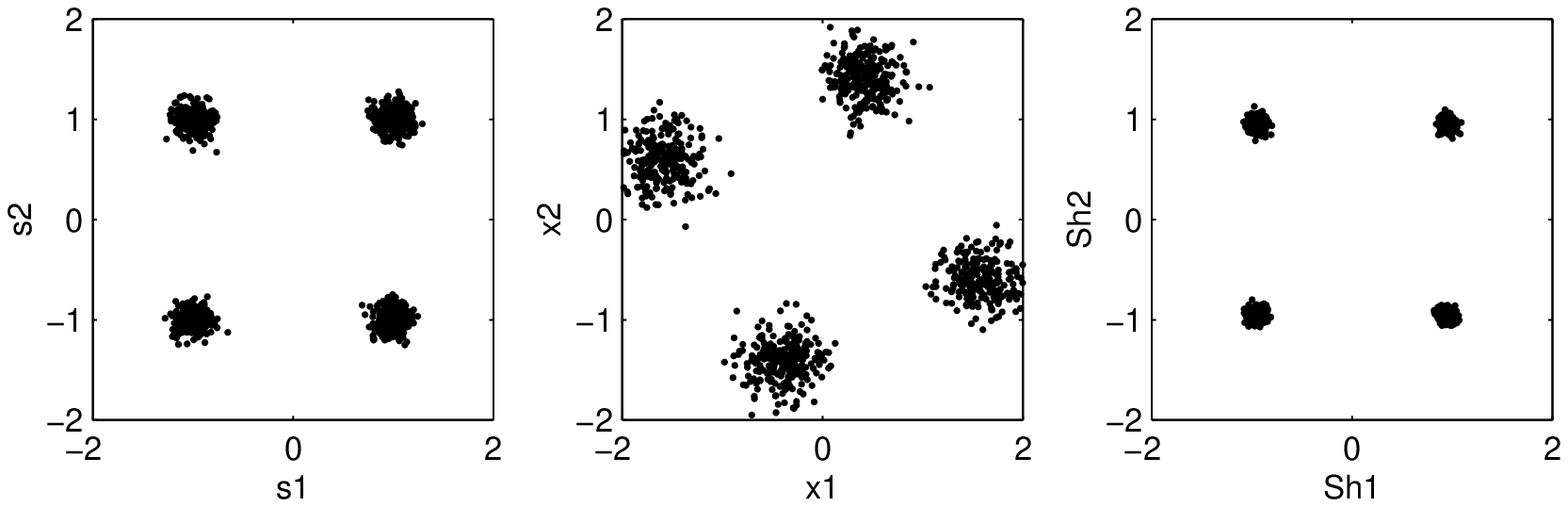} \\
Figure 5- Separation results with $SNR=15\,dB$: Phase space distribution of sources,\\
mixed signals and separated sources.
\etabu

\bigskip
\noindent\btabu{@{}c@{}}
\includegraphics[width=\textwidth,height=60mm]{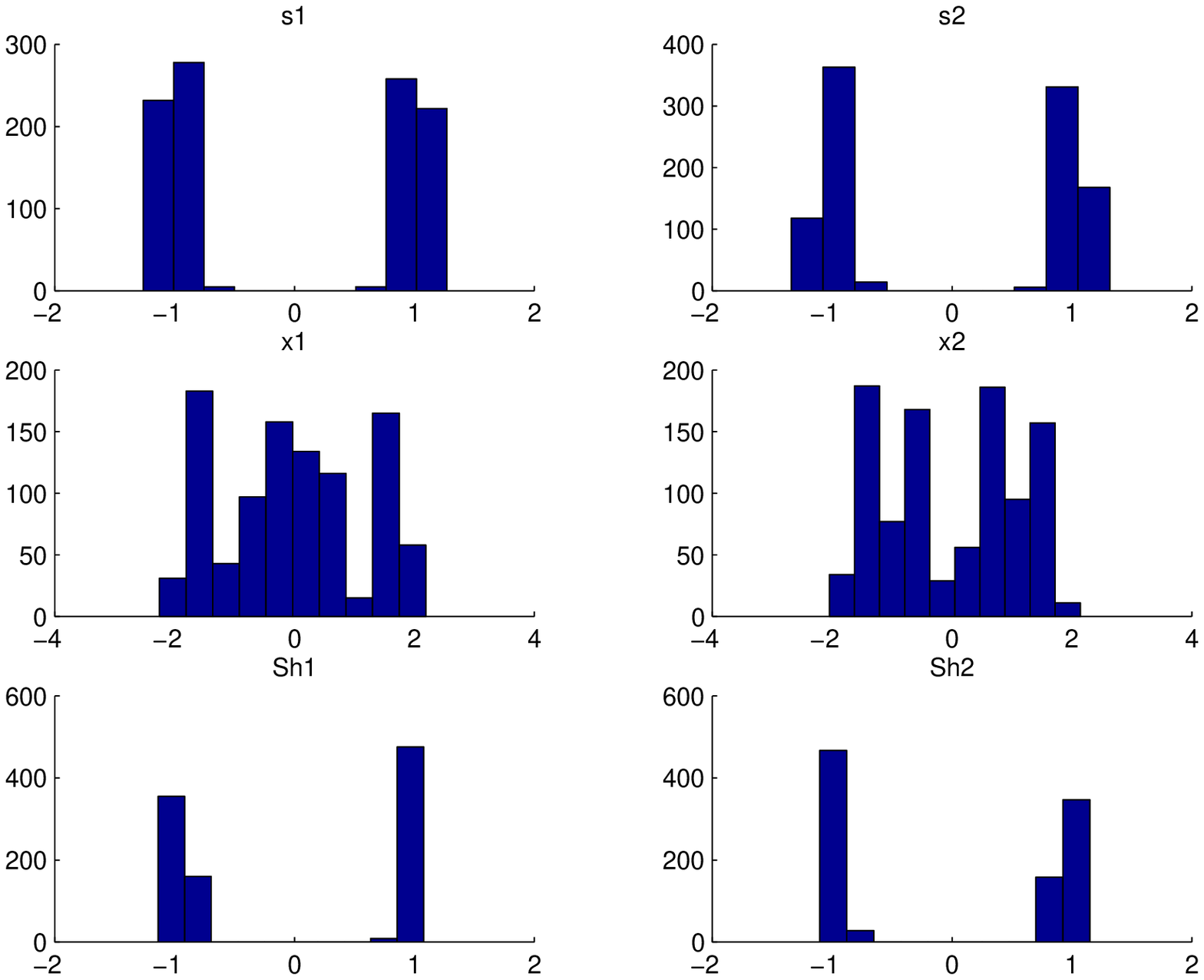}\\~\\
Figure 6- Separation results with $SNR=15\,dB$: Histograms of sources,\\
mixed signals and separated sources.
\etabu

\bigskip
\noindent\btabu{@{}c@{}c@{}}
\includegraphics[width=\textwidth/2,height=50mm]{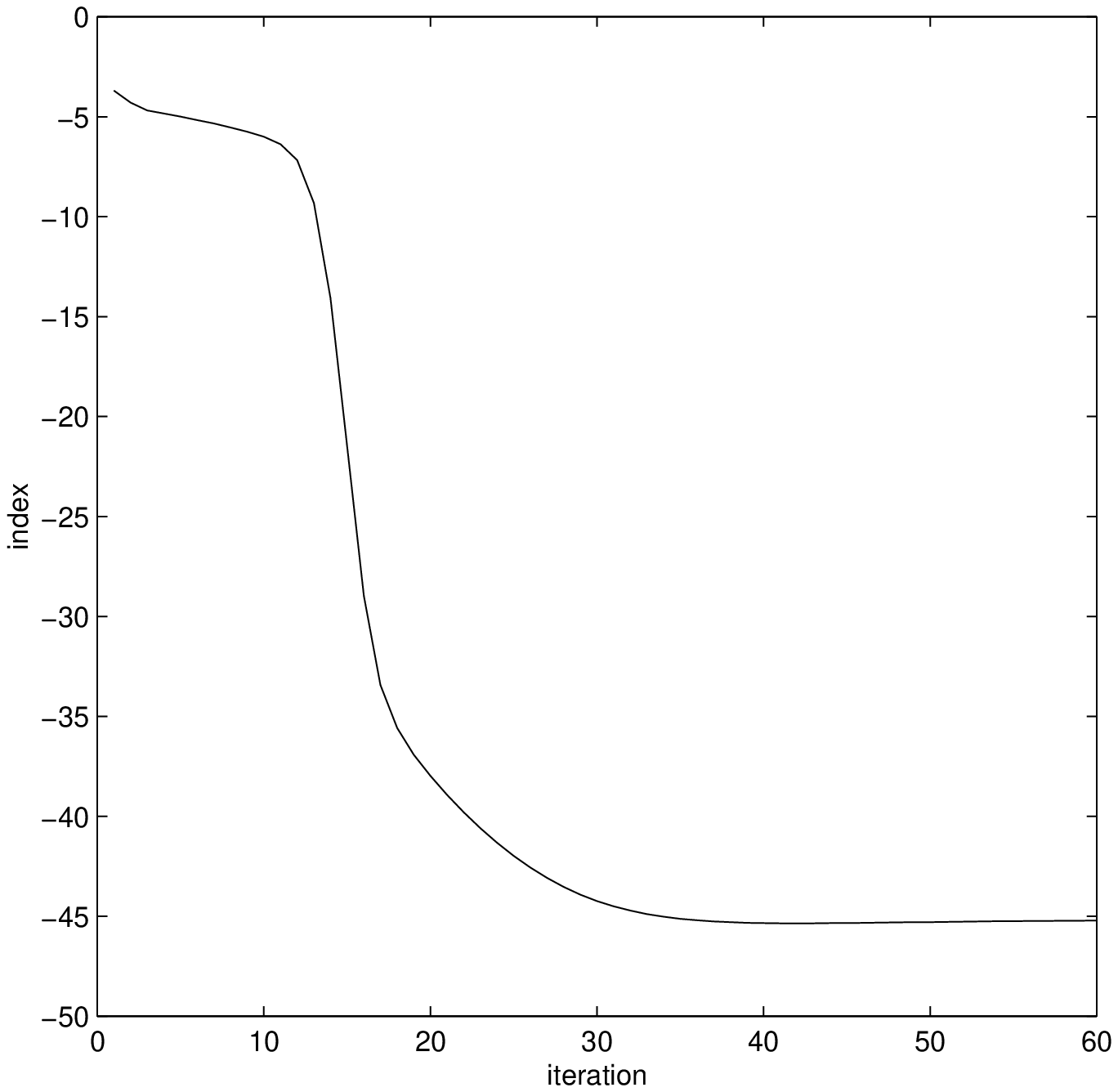}&
\includegraphics[width=\textwidth/2,height=50mm]{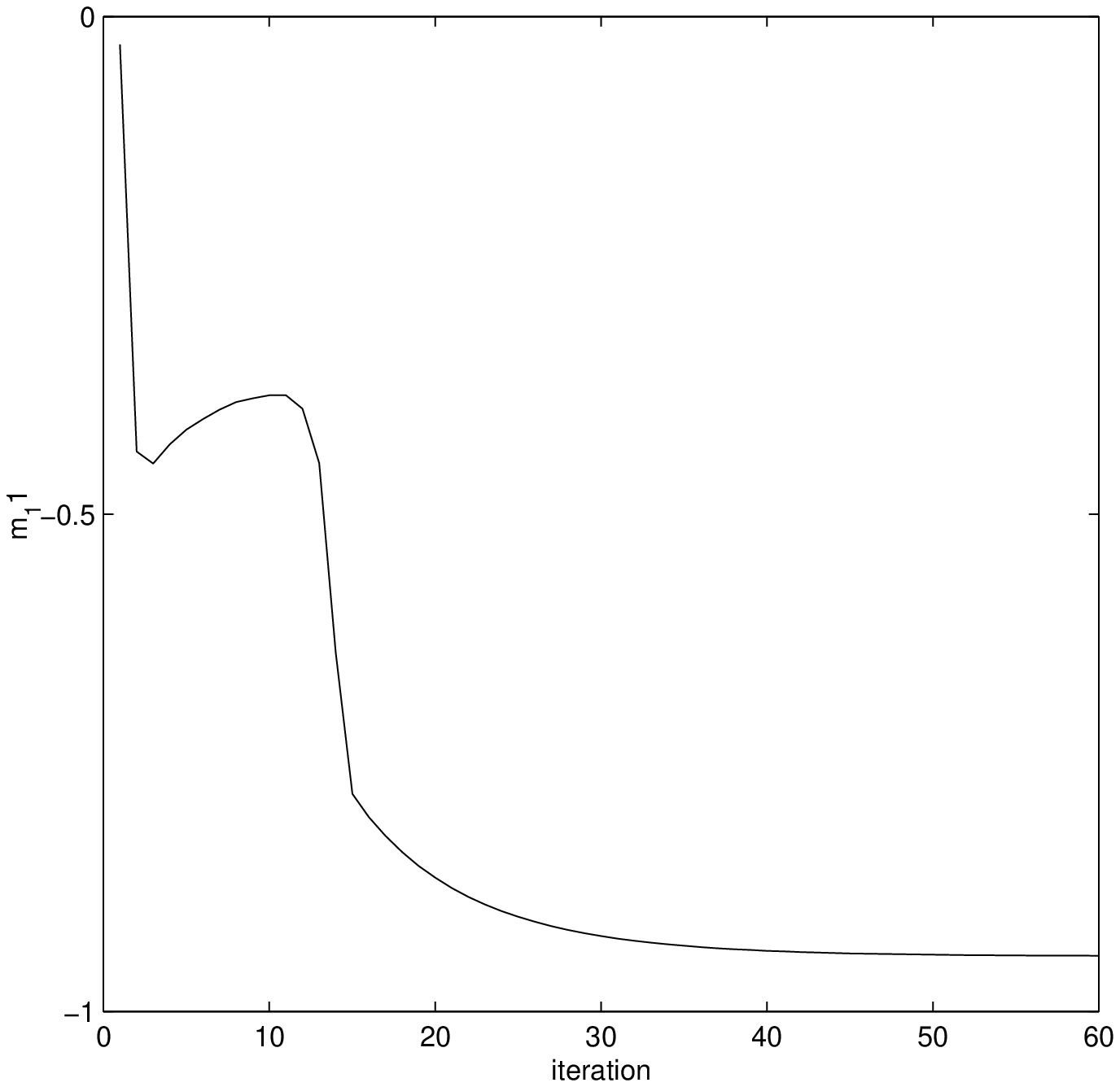}
\\~\\
{\small Figure 7-a- Evolution of index through iterations}& 
{\small Figure 7-b- Identification of $m_{11}$}
\\~\\
\includegraphics[width=\textwidth/2,height=50mm]{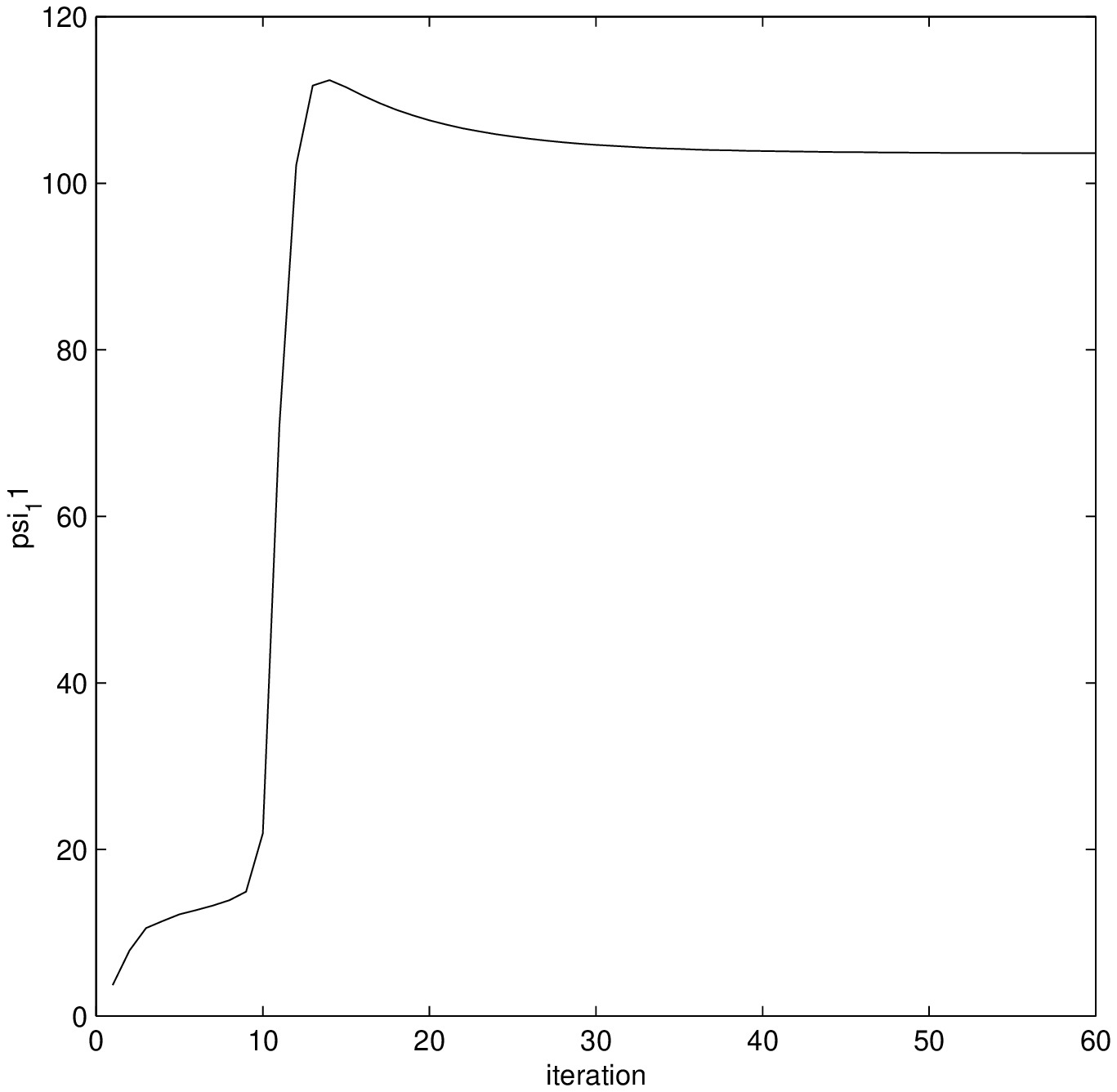}&
\includegraphics[width=\textwidth/2,height=50mm]{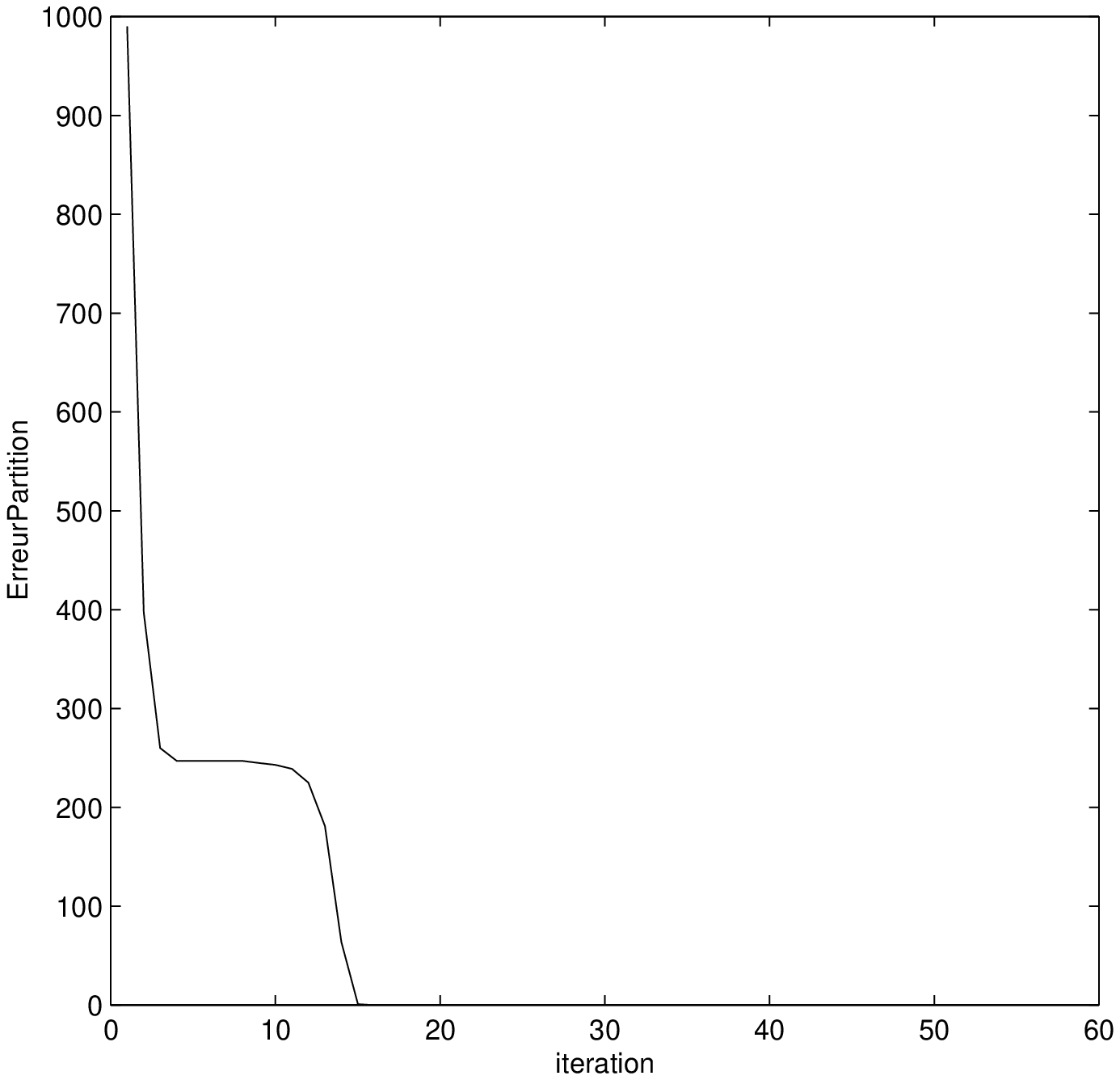}\\~\\
{\small Figure 7-c- Identification of $\psi_{11}$}& 
{\small Figure 7-d- Evolution of classification error}
\etabu

\bigskip
Thus, a joint estimation of sources, mixing matrix and hyperparameters is performed successfully with a JMAP algorithm. The EM algorithm was used in \cite{Bermond00} to solve source separation problem in a maximum likelihood context. We  now  use the EM algorithm in a Bayesian approach  to take into account of our \aprio information on the mixing matrix.

\section{\textit{Penalized EM}} 
The EM algorithm has been used extensively in data analysis to find the maximum likelihood estimation of a set of parameters from given data \cite{Redner84}. Considering both the mixing matrix $\Ab$ and hyperparameters $\thetab$, at the same level, being unknown parameters and complete data $\xb_{1..T}$ and $\sb_{1..T}$. Complete data means jointly observed data $\xb_{1..T}$ and unobserved data $\sb_{1..T}$. The EM algorithm is executed in two steps: ({\bf{i}}) E-step (expectation) consists in forming the logarithm of the joint distribution of observed data $\xb$ and hidden data $\sb$ conditionally to parameters $\Ab$ and $\thetab$ and then compute its expectation conditionally to $\xb$ and estimated parameters  $\Ab^{'}$ and $\thetab^{'}$ (evaluated in the previous iteration), ({\bf{ii}}) M-step (maximization) consists of the maximization of the obtained functional with respect to the parameters $\Ab$ and $\thetab$:
\begin{enumerate}
\item E-step : 
\beq
Q \left( \Ab,\,\thetab\,|\,\Ab',\,\thetab' \right) = E_{\xb,\sb}\left[  \log\,p(\xb,\,\sb\,|\,\Ab,\,\thetab) \,|\,\xb,\,\Ab',\,\thetab' \right]
\eeq
\item M-step : 
\beq
\left( \wh{\Ab},\,\wh{\thetab} \right) = \argmax_{\left( \Ab,\,\thetab\right)}{\{Q \left( \Ab,\,\thetab\,|\,\Ab',\,\thetab' \right)\}}
\eeq
\end{enumerate}
Recently, in \cite{Bermond00}, an EM algorithm has been used in source separation with mixture of Gaussians as sources prior. In this work, we show that:
\begin{enumerate}
\item This algorithm fails in estimating variances of Gaussian mixture. We proved that this is because the degeneracy of the estimated variance to zero. 
\item The computational cost of this algorithm is very high.
\item The algorithm is very sensitive to initial conditions.
\item In \cite{Bermond00}, there's neither an \aprio distribution on the mixing matrix $\Ab$ or on the hyperparameters $\thetab$.
\end{enumerate}  

Here, we propose to extend this algorithm in two ways by:
\begin{enumerate}
\item Introducing an \aprio distribution for $\thetab$ to eliminate degeneracy and an \aprio distribution for $\Ab$ to express our previous knowledge on the mixing matrix. 
\item Taking advantage of our hierarchical model and the idea of classification to reduce the computational cost.
\end{enumerate}
To distinguish the proposed algorithm from the one proposed in \cite{Bermond00}, we call this algorithm the \textit{Penalized EM}. The two steps become:

\begin{enumerate}
\item E-step : 
\beq 
Q \left( \Ab,\,\thetab\,|\,\Ab',\,\thetab' \right) = E_{\xb,\sb}\left[  \log\,p(\xb,\,\sb\,|\,\Ab,\,\thetab)+\log\,p(\Ab)+\log\,p(\thetab) \,|\,\xb,\,\Ab',\,\thetab' \right]
\eeq

\item M-step : 
\beq 
\left( \wh{\Ab},\,\wh{\thetab} \right) = \argmax_{\left( \Ab,\,\thetab\right)}{Q \left( \Ab,\,\thetab\,|\,\Ab',\,\thetab' \right)}
\eeq
\end{enumerate}   

The joint distribution is factorized as: $p(\xb,\,\sb,\,\Ab,\,\thetab)\,=\,p(\xb\,|\,\Ab,\,\sb)\,p(\Ab)\,p(\sb\,|\,\thetab)\,p(\thetab)$. We can remark that $p(\xb,\,\sb,\,\Ab,\,\thetab)$ as a function of $(\Ab,\,\thetab)$ is separable in $\Ab$ and $\thetab$. Consequently, the functional is  separated into two factors: one representing an $\Ab$ functional and the other representing a $\thetab$ functional:
\beq
Q \left( \Ab,\,\thetab\,|\,\Ab',\,\thetab' \right) = Q_a \left( \Ab\,|\,\Ab',\,\thetab' \right)+Q_h \left( \thetab\,|\,\Ab',\,\thetab' \right)
\eeq
with:
\beq \left\{ \barr{ccc}
Q_a \left( \Ab\,|\,\Ab',\,\thetab' \right) &= & E\left[  \log\,p(\xb\,|\,\Ab,\,\sb)+\log\,p(\Ab) \,|\,\xb,\,\Ab',\,\thetab' \right] \\
Q_h \left( \thetab\,|\,\Ab',\,\thetab' \right) &= & E\left[  \log\,p(\sb\,|\,\thetab)+\log\,p(\thetab) \,|\,\xb,\,\Ab',\,\thetab' \right] \\
\earr \right.
\eeq 

\subsection*{- Maximisation with respect to $\Ab$}
The functional $Q_a$ is:
\beq
Q_a=\frac{-1}{2\,\sigma_{\epsilon}^2} \sum_{t=1}^{T} E\left[ \left( \xb(t)-\Ab\,\sb(t)\right)^{T}\left( \xb(t)-\Ab\,\sb(t)\right)|\,\xb,\,\Ab',\,\thetab' \right]+ \log\,p(\Ab).
\eeq
The gradient of this expression with respect to the elements of $\Ab$ is:

\beq \label{gradient}
 \frac{\partial{Q_a}}{\partial{\Ab_{i,j}}}=\frac{T}{\sigma_{\epsilon}^2}\left( \wh{\Rb}_{xs}-\Ab\,\wh{\Rb}_{ss} \right)_{i,j}-\frac{1}{\sigma_{aij}^2} \left( \Ab_{i,j}-\Mb_{i,j} \right).
\eeq

where:

\beq \left\{ \barr{ccc}
\wh{\Rb}_{xs} &=& \frac{1}{T}\sum_{t=1}^{T} E\left[  \xb(t)\, \sb(t)^{T}|\,\xb,\,\Ab',\,\thetab' \right]\\
\wh{\Rb}_{ss} &=& \frac{1}{T}\sum_{t=1}^{T} E\left[  \sb(t)\, \sb(t)^{T}|\,\xb,\,\Ab',\,\thetab' \right]
\earr \right.
\eeq

Evaluation  of $\wh{\Rb}_{xs}$ and $\wh{\Rb}_{ss}$ requires the computation of the expectations of $\xb(t)\, \sb(t)^{T}$ and $\sb(t)\, \sb(t)^{T}$. The main computational cost is due to the fact that the expectation of any function $ f\left( \sb \right)$ is given by:
\beq\label{somme}
E\left[\,  f\left( \sb \right)\,|\,\xb,\,\Ab',\,\thetab' \right]=\sum_{\zb' \,\in\, \prod_{i=1}^{n}\Zc_i}E\left[\,  f\left( \sb \right)\,|\,\xb,\zb=\zb',\,\Ab',\,\thetab' \right]\, p(\zb'\,|\,\xb,\,\Ab',\,\thetab').
\eeq

\noindent which involves a sum of $\prod_{j=1}^{n}q\left(j\right)$ terms corresponding to the whole combinations of labels. One way to obtain an approximate but fast estimate of this expression is to limit the summation to only one term corresponding to the MAP estimate of $\zb$:
\beq
E\left[\,  f\left( \sb \right)\,|\,\xb,\,\Ab',\,\thetab' \right]=E\left[\,  f\left( \sb \right)\,|\,\xb,\zb=\wh{\zb}^{MAP},\,\Ab',\,\thetab' \right].
\eeq
Then, given estimated labels $\zb_{1..T}$, the source $\sb(t)$  \apost law is Normal with  mean $\thetab_{xz}$ and variance $\Vb_{xz}$ given by (\ref{moyenne}) and (\ref{variance2}).

The source estimate is then $\thetab_{xz}$. $\wh{R}_{xs}$ and $\wh{R}_{ss}$ become:
\beq
\wh{R}_{xs} = \frac{1}{T}\sum_{t=1}^{T} \xb(t)\, \wh{\sb}(t)^{T}
\eeq
and
\beq
\wh{R}_{ss} = \frac{1}{T}\sum_{t=1}^{T} \wh{\sb}(t)\, \wh{\sb}(t)^{T}+\frac{1}{T}\sum_{t=1}^{T} (\Ab^{t}\Rb_n^{-1}\Ab+ \bf{\Gamma}_z^{-1})^{-1} 
\eeq

When $\Sb_{1..T}$ estimated and  using the matrix operations defined in section II and cancelling the gradient (\ref{gradient}) to zero, we obtain the expression of the estimate of $\Ab$:
\beq\label{melange}
\Abh^{k+1}=Mat\left(\left[ \Lambdab+ T\,\wh{R}_{ss}^{*} \right]^{-1}\left[ \Lambdab Vect(M)+T\,Vect\left( \wh{R}_{xs}\right) \right] \right)
\eeq

\subsection*{- Maximisation with respect to $\thetab$}

With a uniform \aprio for the means, maximisation of $Q_h$ with respect to $m_{jz}$ gives :
\beq\label{hyper1}
\wh{m}_{jz}=\frac{\sum_{t=1}^{T}\theta_{jz}(t)\,p(z(t)\,|\,\xb,\,\Ab',\,\thetab')}{\sum_{t=1}^{T}p(z(t)\,|\,\xb,\,\Ab',\,\thetab')}
\eeq

With an Inverted Gamma prior $\Gc\left(\alpha,\,\beta\right)$ ($\alpha > 0$ et $\beta > 1$) for the variances, the maximisation of $Q_h$ with respect to $\sigma_{jz}$ gives:

\beq\label{hyper2}
\wh{\sigma}_{jz}=\frac{2\,\beta+\sum_{t=1}^{T}\left(V_{jz}+\theta_{jz}^2-2\,\wh{m}_{jz}\theta_{jz}+\wh{m}_{jz}^2\right)p(z(t)\,|\,\xb,\,\Ab',\,\thetab')}{\sum_{t=1}^{T}p(z(t)\,|\,\xb,\,\Ab',\,\thetab')+2\,\left(\alpha-1\right)}
\eeq

\subsection*{Summary of the Penalized EM algorithm}
Based on the preceeding equations, we propose the following algorithm to estimate sources and parameters using the following five steps:
\begin{enumerate}
\item Estimate the hyperparameters according to (\ref{hyper1}) and (\ref{hyper2}).
\item Update of data classification by estimating $\wh{\zb}_{1..T}^{MAP}$.
\item Given this classification, sources estimate is the mean of the Gaussian \apost law (\ref{mean}).
\item Update of data classification.
\item Estimate the mixing matrix $\Ab$ according to the re-estimation equation (\ref{melange}).
\end{enumerate}

\section{Comparison with JMAP algorithm and its sensitivity to initial conditions}
The Penalized EM algorithm has an optimization cost  approximately $2$ times higher, per sample, than the JMAP algorithm. However, both algorithms have a reasonable computational  complexity, linearly increasing with the number of samples.
 Sensitivity to initial conditions is inherent to the EM-algorithm even to the penalized version. In order to illustrate this fact, we simulated the algorithm with the same parameters as in section IV. Note that initial conditions for hyperparameters are $\psi^{(0)}=\left(\barr{cc}1 & 1 \\ 1 & 1\earr\right)$ and  $m^{(0)}=\left(\barr{cc}0 & 0 \\ 0 & 0\earr\right)$. However, the Penalized EM algorithm fails in separating sources (see figure 11). We note then that JMAP algorithm is more robust to initial conditions.

\bigskip
\noindent\btabu{@{}c@{}}
\includegraphics[width=\textwidth,height=40mm]{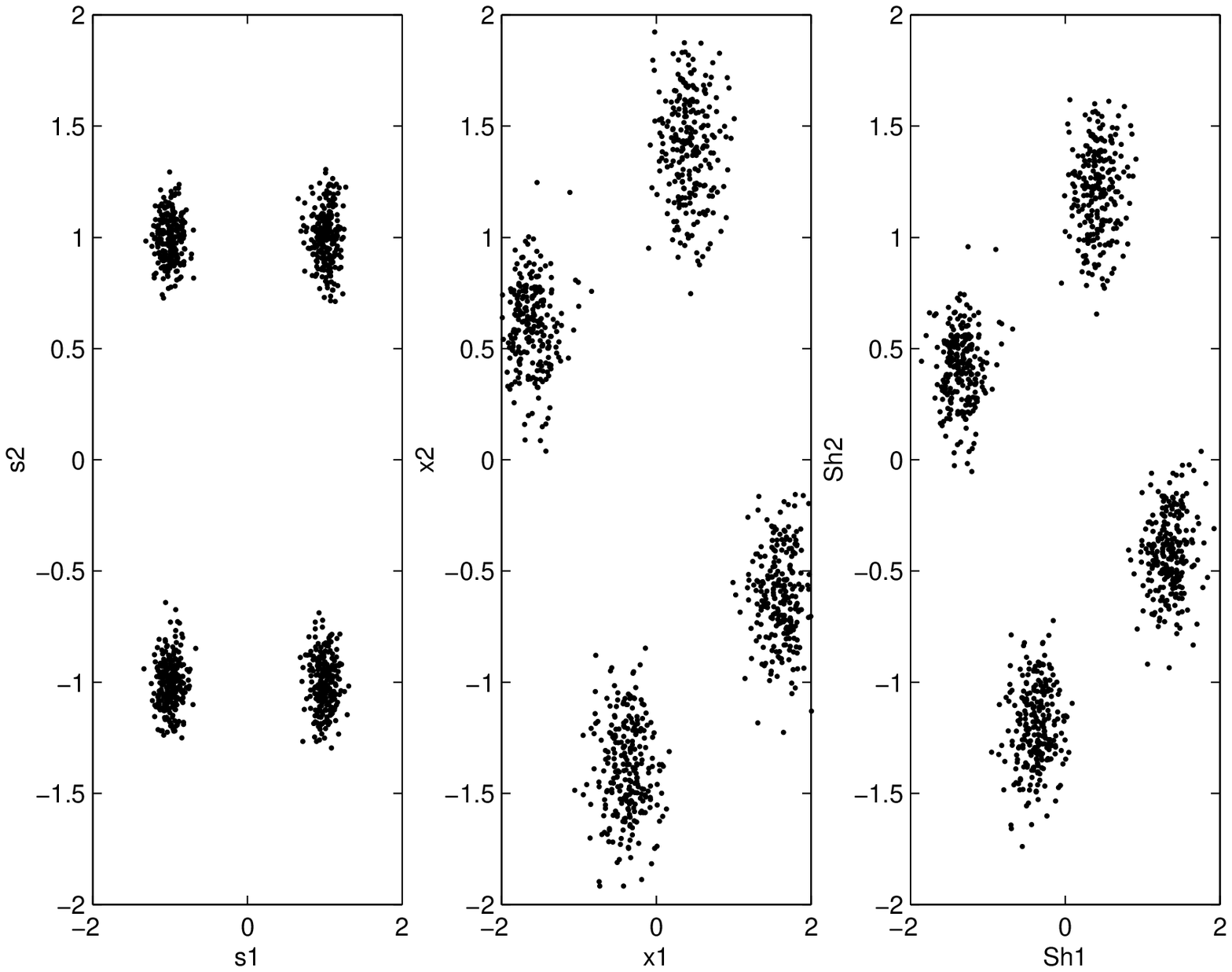}\\
\hspace*{0.3cm}(a)\hspace*{4.5cm}(b)\hspace*{4.5cm}(c)\\~\\
Figure 11- Results of separation with the Penalized EM algorithm: \\
 (a) Phase space distribution of sources,\\ 
 (b) mixed signals and (c) separated sources
\etabu

We modified the initial condition to have means:  
$m^{(0)}=\left(\barr{cc}
-0.5 & 0.5 \\ 
-0.5 & 0.5
\earr\right)$. 
We noted, in this case,  the convergence of the Penalized EM algorithm 
to the correct solution.  
Figures $12$-$16$ illustrate the separation results:

\bigskip
\noindent\btabu{@{}c@{}}
\includegraphics[width=\textwidth,height=40mm]{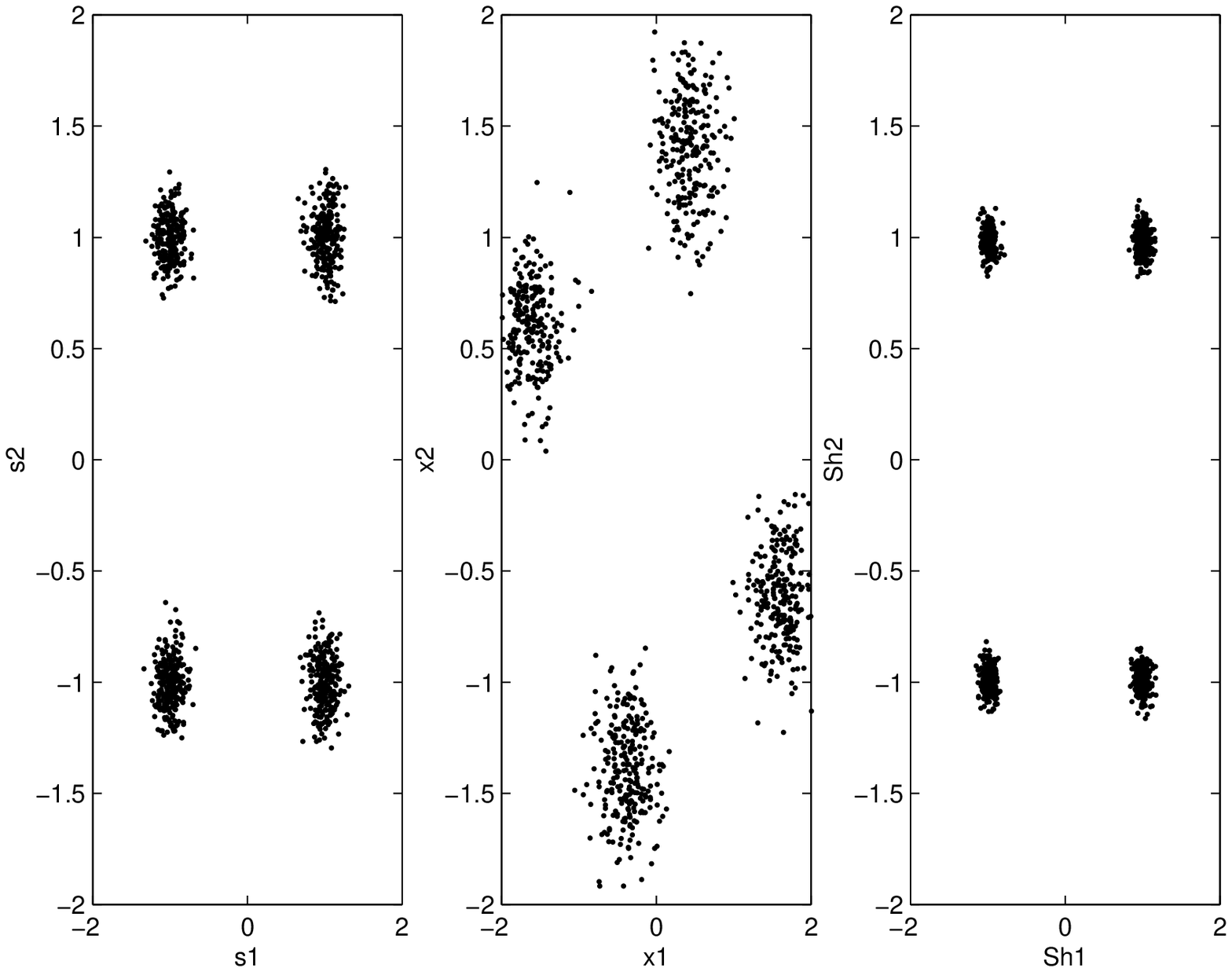}\\
\hspace*{0.3cm}(a)\hspace*{4.5cm}(b)\hspace*{4.5cm}(c)\\~\\
Figure 12- Results of separation with the Penalized EM algorithm: \\
 (a) Phase space distribution of sources,\\ 
 (b) mixed signals and (c) separated sources
\etabu

\bigskip
\noindent\btabu{@{}c@{}c@{}}
\includegraphics[width=\textwidth/2,height=40mm]{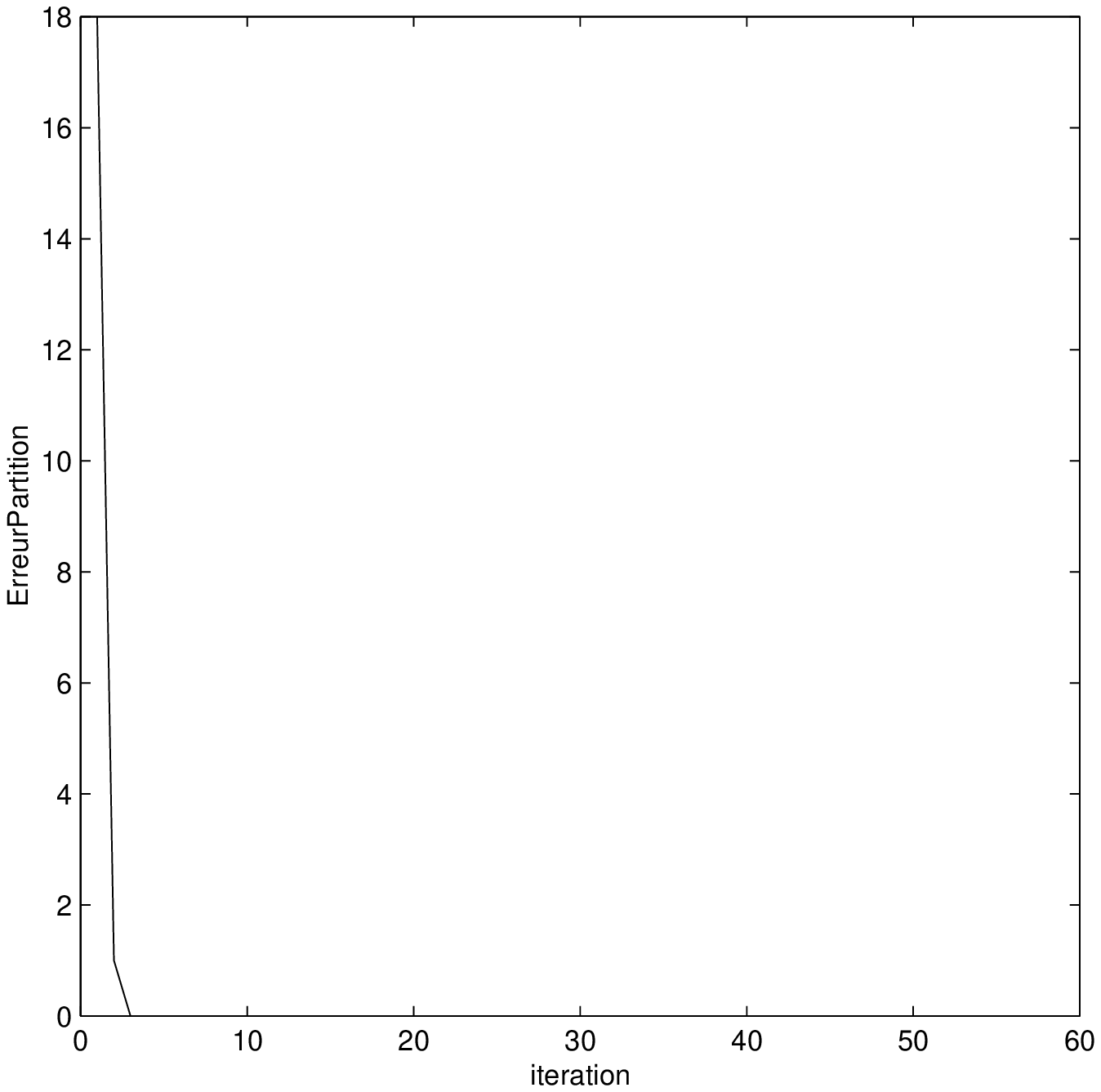}&
\includegraphics[width=\textwidth/2,height=40mm]{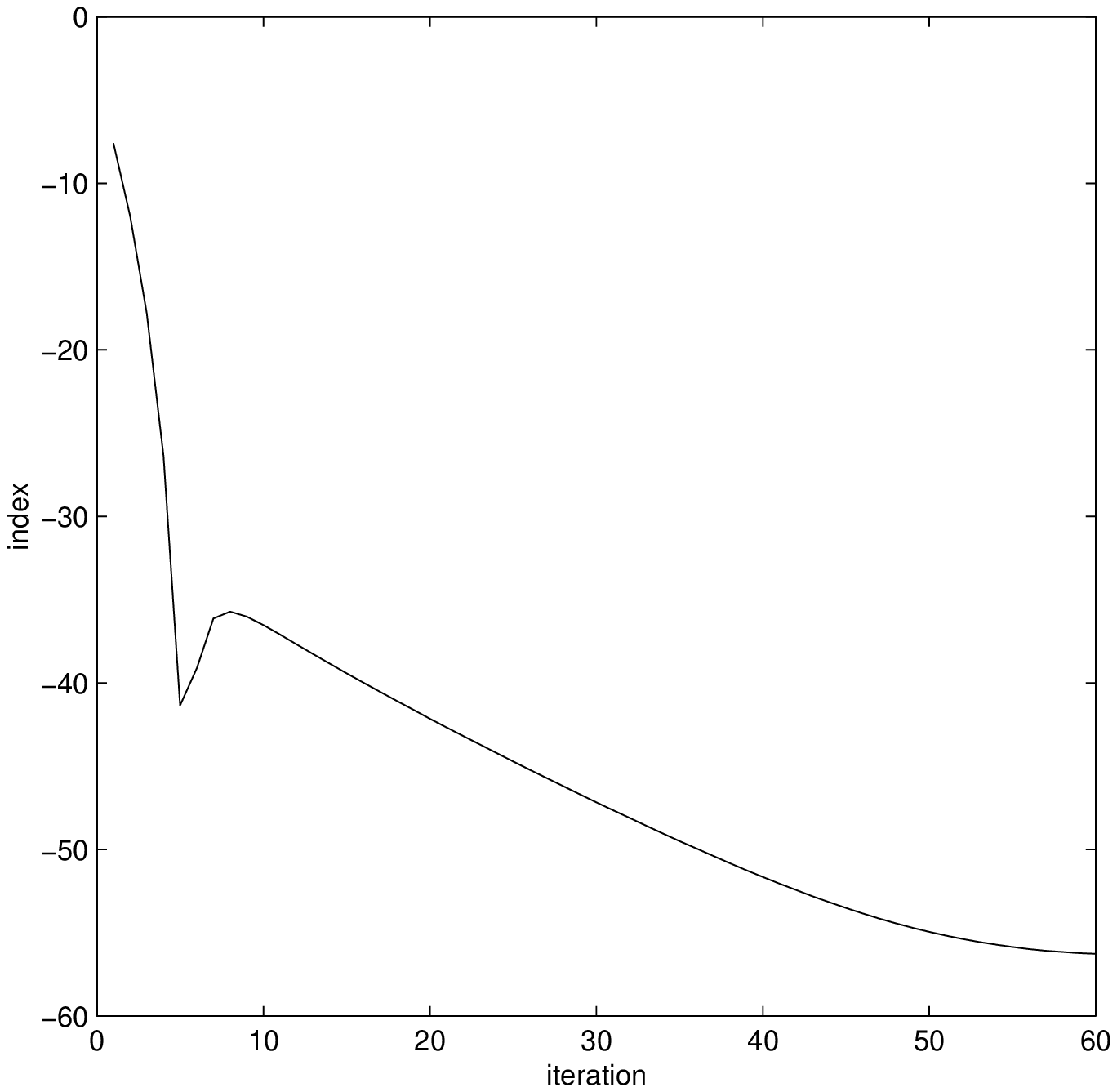}\\
Figure 13- Evolution of classification error&
Figure 14- Evolution of index
\etabu

\bigskip
\noindent\btabu{@{}c@{}c@{}}
\includegraphics[width=\textwidth/2,height=40mm]{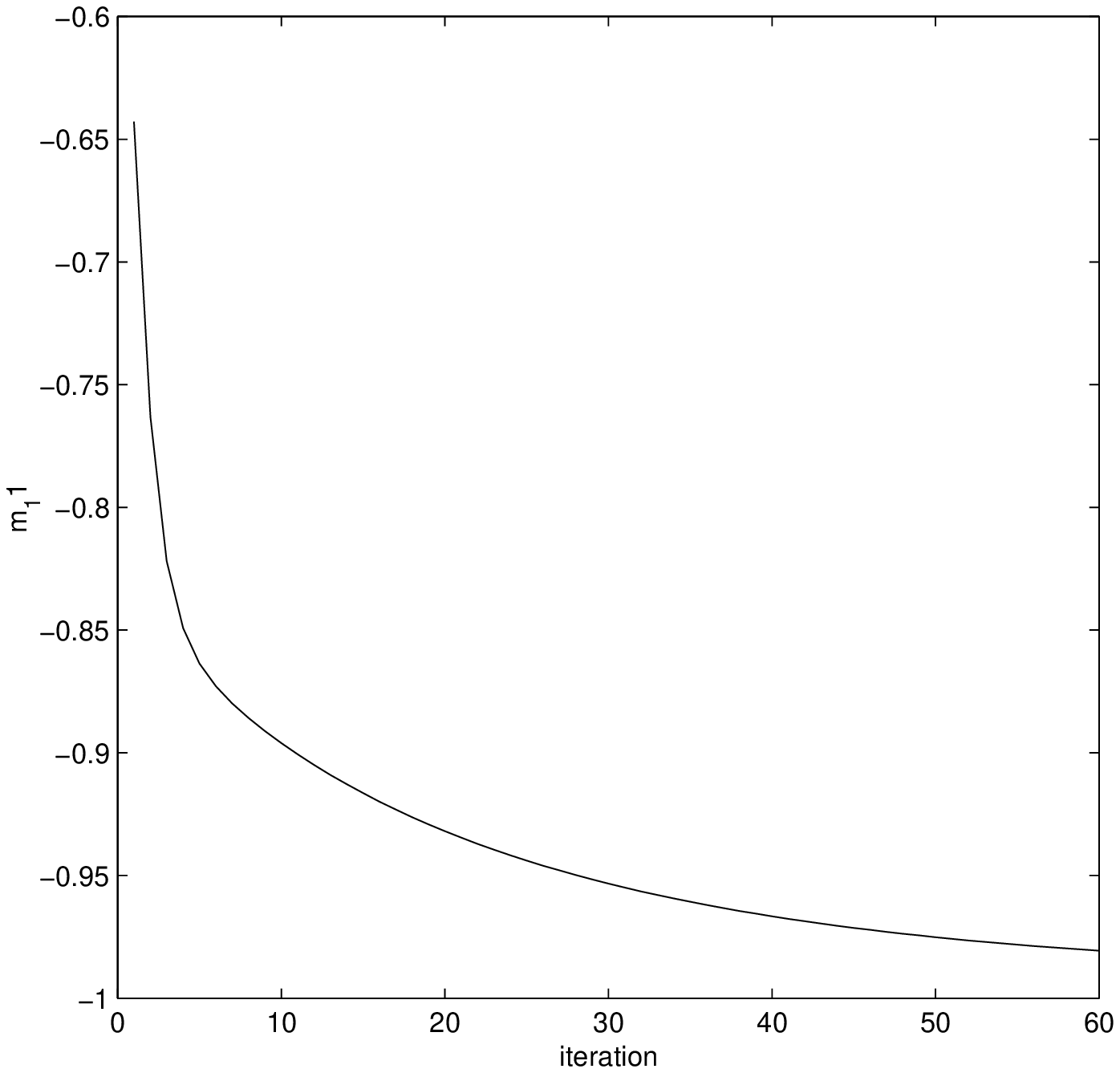}&
\includegraphics[width=\textwidth/2,height=40mm]{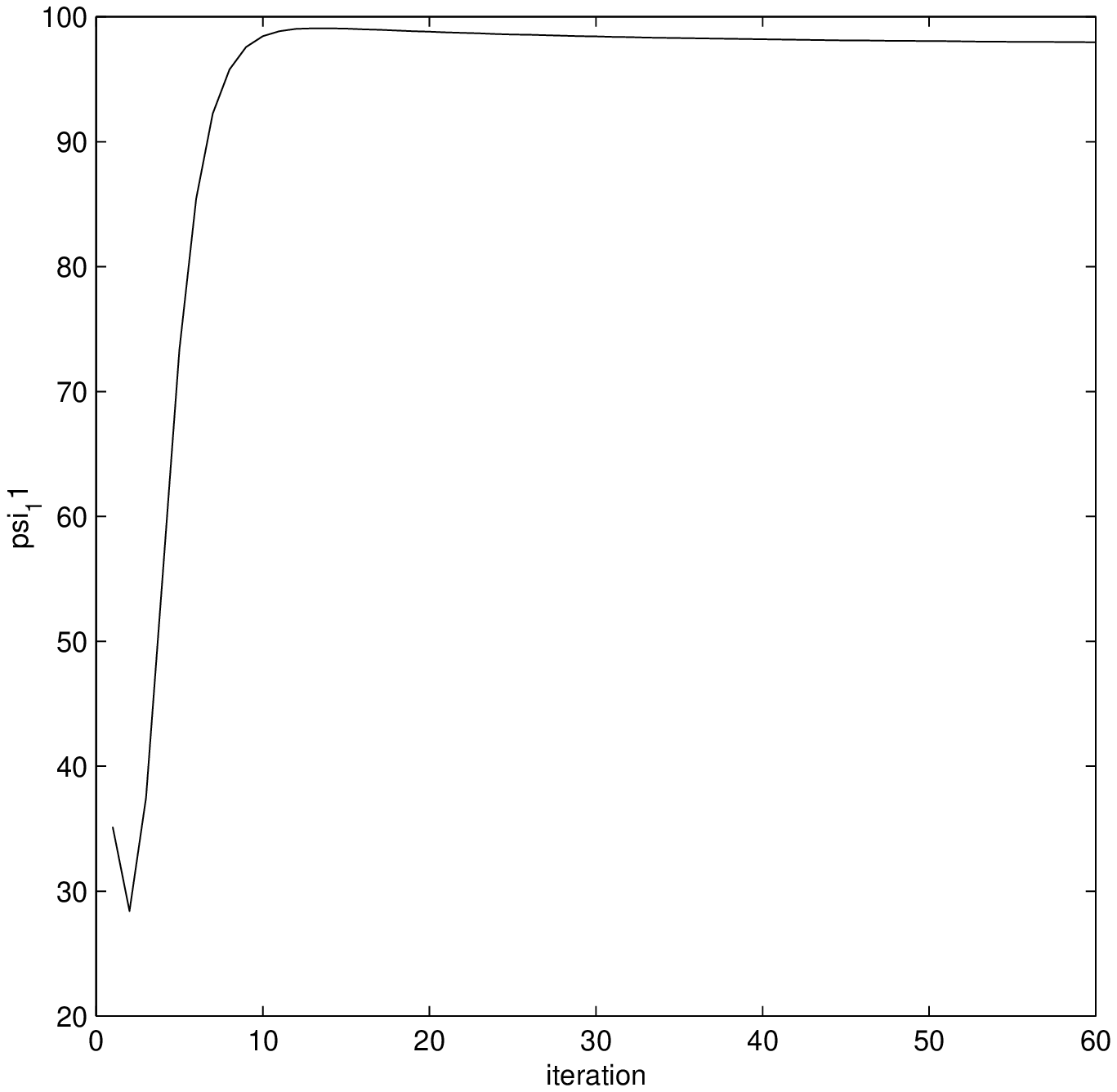}\\
Figure 15- Identification of $m_{11}$&
Figure 16- Identification of $\psi_{11}$
\etabu

\section{Conclusion}
We have proposed solutions to source separation problem using a Bayesian framework. Specific aspects of the described approach include:
\bit
\item Taking account of errors on model and measurements.
\item Introduction of \aprio distribution for the mixing matrix and hyperparameters. This was motivated by two different reasons: Mixing matrix prior should exploit  previous information and variances prior should  regularize the log-posterior objective function.
\eit
We then consider the problem in terms of a  mixture of Gaussian priors to develop a hierarchical strategy for source estimation. This same interpretation leads us to classify data before estimating hyperparameters and to reduce computational cost in the case of the proposed Penalized EM algorithm.
{\small 

\bibliographystyle{maxent95}

}
\edoc